\documentclass[]{aa}  

\usepackage{graphicx}
\usepackage{txfonts}
\usepackage{amsmath}
\usepackage{natbib}
\setcitestyle{authoryear,open={(},close={)}} 

\begin{document}

   \title{Coronal energy release by MHD avalanches}

   \subtitle{Effects on a structured, active region, multi-threaded coronal loop}

   \author{G. Cozzo\inst{1}
          \and
          J. Reid \inst{2}
          \and
          P. Pagano \inst{1,3}
          \and
          F. Reale \inst{1,3}
          \and
          A.~W. Hood \inst{2}
          }

   \institute{
            Dipartimento di Fisica \& Chimica, Università di Palermo, Piazza del Parlamento 1, I-90134 Palermo, Italy
            \email{gabriele.cozzo@unipa.it}
            \and School of Mathematics and Statistics, University of St Andrews, St Andrews, Fife, KY16 9SS, UK
            \email{jr93@st-andrews.ac.uk}
            \and INAF-Osservatorio Astronomico di Palermo, Piazza del Parlamento 1, I-90134 Palermo, Italy
            \email{paolo.pagano@inaf.it}          
            }
   \date{Received \dots ; accepted \dots}

 
  \abstract
   {A possible key element for large-scale energy release in the solar corona is an MHD kink instability in a single twisted magnetic flux tube. An initial helical current sheet progressively fragments in a turbulent way into smaller-scale sheets. Dissipation of these sheets is similar to a nanoflare storm. Since the loop expands in the radial direction during the relaxation process, an unstable loop can disrupt nearby stable loops and trigger an MHD avalanche.}
   {Exploratory investigations have been conducted in previous works with relatively simplified loop configurations.
   Here, we address a more realistic environment that comprehensively accounts for most of the physical effects involved in a stratified atmosphere, typical of an active region.
   The question is whether the avalanche process will be triggered, with what timescales, and how it will develop, as compared with the original, simpler approach.}
   {Three-dimensional (3D) MHD simulations describe the interaction of magnetic flux tubes, which have a stratified atmosphere, including chromospheric layers, the thin transition region to the corona, and the related transition from high-$\beta$ to low-$\beta$ regions. The model also includes the effects of thermal conduction and of optically thin radiation.}
   {Our simulations address the case where one flux tube among a few is twisted at the footpoints faster than its neighbours. We show that this flux tube becomes kink unstable first, in conditions in agreement with those predicted by analytical models. It rapidly involves nearby stable tubes, instigating significant magnetic reconnection and dissipation of energy as heat. The heating determines, in turn, the development of chromospheric evaporation, while the temperature rises up to about $10^7\,\mathrm{K}$, close to microflares observations.}
   {This work therefore confirms, in more realistic conditions, that avalanches are a viable mechanism for the storing and release of magnetic energy in plasma confined in closed coronal loops, as a result of photospheric motions.}


   \maketitle
%
\section{Introduction} \label{sec:introduction}
The magnetic activity in the solar corona consists of a wide range of events, reflecting the dynamic nature of the environment.
The role of the magnetic field as the dominant reservoir of energy that may heat the corona to millions of kelvin is now widely accepted.
Moreover, it is now clear that the \textit{coronal heating problem} must address the whole solar atmosphere as a highly coupled system \citep{parnell2012contemporary}: the solar corona must not be treated as an isolated environment, but as an energetically open system with continuous interactions with the underlying layers (i.e. the chromosphere and photosphere).
On the other hand, the nature of the mechanisms that might lead the magnetic field to supply energy to the corona is still highly debated. Although several processes have been proposed in recent decades, it is still unclear how they interact to produce the complex behaviour of the solar atmosphere. For instance, photospheric observations show a very clumpy magnetic field, organized into clusters of elemental flux tubes \citep{gomez1993normal}.
Above that, the bright corona consists of arch-like magnetic structures which connect regions with different polarity: coronal loops \citep{vaiana1973identification, reale2014coronal}. Loops are, in turn, structured into thinner magnetic strands which reflect the underlying granular pattern. 
Tangling and twisting of the coronal magnetic strands cannot be avoided, according to photospheric observations. The field must therefore reconnect in order to prevent an infinite build-up of stress. This inevitably produces plasma heating \citep{parker1972topological, klimchuk2015key}. Investigating coronal loops is of fundamental importance in order to understand key physical aspects of several heating mechanisms based on magnetic stress.

Coronal loops are generally organised into clusters of thin, twisted threads (also called \textit{strands}) following the same collective behaviour. Nevertheless, as coronal loops commonly exhibit strong magnetic fields of the order of $10\,\mathrm{G}$ or more \citep{yang2020global,long2017measuring,Brooks2021}, and the coronal plasma nearly ideal, transport of matter across the field is strongly inhibited. In other words, the magnetic field prevents the plasma from directly diffusing outside of the flux tubes. Moreover, since magnetic forces are much stronger than gravity in the corona, the latter will effectively act only along the field lines. Tenuous plasma is strongly funnelled along the field lines and also thermally isolated from the surroundings \citep{rosner1978dynamics, vesecky1979numerical}.
Coronal loops are anchored to the underlying chromosphere and, a little further down, to the photospheric layer, where the plasma $\beta$ parameter exceeds unity by a few orders of magnitude. There, the so-called loop \textit{foot-points} are dragged by photospheric plasma motion, which, in turn, might be highly irregular and turbulent. The typical strength of the magnetic field in the photosphere is found to be hundreds of Gauss in active regions and sunspots \citep{ishikawa2021mapping}. 
Higher in the corona, the plasma pressure decreases, the flux tube progressively expands, and the magnetic field strength decreases, while keeping the magnetic flux conserved.
The greatest expansion rate is expected across the thin transition region separating the chromosphere from the overlying corona \citep{gabriel1976magnetic}. There, the temperature rapidly increases from thousands to millions of kelvin and, consequently, the pressure scale-height increases by several orders of magnitude.

The solar corona may be heated both by dissipation of
stored magnetic stresses (DC heating) and by the damping of waves (AC heating) \citep{parnell2012contemporary, zirker1993coronal}.
In particular, DC heating must involve both storage and impulsive release of magnetic energy. In solar active regions, the energy is presumably stored over timescales longer than an Alfvén time. A reasonable general assumption is that the magnetic field evolves quasi-statically through a sequence of equilibria, slowly changing because the coronal footpoints are rigidly line-tied to the low-beta photospheric plasma. As the coronal field moves through these equilibria, energy is injected by the motions, and is then stored.
Observations and numerical experiments provide evidence that the evolution of coronal loops is strongly influenced by the photospheric motions \citep{chen2021transient}. The coronal magnetic field must be driven toward a stressed state, which will be a non-potential configuration.
For instance, footpoint rotation may lead the magnetic structure to twist and gain magnetic energy. While magnetic energy is stored, the flux tube could potentially be subject to strong stresses that may eventually trigger fast magnetohydrodynamic instabilities, such as the kink instability \citep[][]{hood2009coronal} or the tearing mode instability \citep{del2016ideal}, or lead to long-lasting Ohmic heating \citep{klimchuk2006solving}. The details of this conversion, for instance whether continuous or by sequences of pulses (nanoflares), are still under investigation. Another outstanding issue is the spatial distribution of the heating, which may reveal a filamentary structure on coronal loops.

Heating and brightening of coronal loops may occur as a \textit{storm} of impulsive events \citep{klimchuk2009coronal, viall2011patterns}. Such heat pulses may be driven by multiple localized instances of the magnetic field relaxing.
The irregular photospheric motion, as well as a large range of magnetohydrodynamic instabilities, may lead the magnetic structure to develop fast reconnection and to produce heat. 
A very compelling body of evidence now supports magnetic reconnection as the key element to start large-scale energy release process, which might dissipate into a background heating \citep{hood2009coronal}.
Apart from individual reconnection events in localized current sheets and neutral points, many sites of reconnection might develop in the complex and dynamic coronal magnetic field. Energy released by several localized reconnection events can be predicted by Taylor's dissipation theory \citep{taylor1974relaxation}, first applied to Reverse Field
Pinch devices in plasma laboratories \citep{taylor1986relaxation} and then extended to the coronal environment \citep{browning1986heating}.
According to Taylor's theory, a turbulent, resistive plasma can rapidly reach a minimum-energy state. During the process, the topology of the magnetic field changes via reconnection, but the magnetic helicity is conserved.

In the corona, the magnetic field might become unstable to resistive modes, as it is slowly forced by photospheric motions to explore a series of non-linear force-free states. In conditions of high magnetic stress the field must reconnect and relax towards a linear force-free state, $\nabla \times \mathbf{B} = \alpha \mathbf{B}$, with uniform $\alpha$ \citep{woltjer1958theorem, heyvaerts1983coronal}.
In particular, magnetic energy is found to be released in the corona throughout a widespread range of events that occur from large (\textit{flares}, $<10^{25}\,\mathrm{J}$) down to medium (\textit{microflares}, $<10^{22}\,\mathrm{J}$) scales \citep[e.g., shown by][]{priest2014magnetohydrodynamics}. 
It has been suggested that the same mechanism, operating on even smaller scales, could be responsible for maintaining the one-million kelvin diffuse corona, through so-called \textit{nanoflare} activity \citep{parker1988nanoflares}.

Undetectable when first proposed \citep{parker1988nanoflares, antolin2021reconnection}, observational evidence of such small events, including nanoflares, has been growing \citep{Mondal2021, Vadawale2021}. 
Solar Orbiter \citep{muller2020solar}’s combination of high-resolution measurements of the photospheric magnetic field with the Polarimetric and Helioseismic Imager \citep[PHI; ][]{PHI}, together with UV and EUV images from the Extreme Ultraviolet Imager \citep[EUI; ][]{EUI} and spectra from the Spectral Imaging of the Coronal Environment \citep[SPICE; ][]{SPICE}, are able to capture localized reconnection events down to even smaller scales, the so-called \textit{campfires} \citep{berghmans2021extreme, zhukov2021stereoscopy}, small and localized brightenings in a quiet Sun region with length scales between $400\,\mathrm{km}$ and $4000\,\mathrm{km}$ and durations between $10\,\mathrm{s}$ and $200\,\mathrm{s}$. These coronal events  are rooted in the magnetic flux concentrations of the chromospheric network. 

A possible trigger mechanism for large-scale energy release (such as solar flares) is the MHD kink instability in a single twisted magnetic flux strand \citep{hood1979kink, hood2009coronal}. 
It typically arises in narrow, strongly twisted magnetic tubes and results in the cross-section of the plasma column moving transversely away from its centre of mass, determining an irreversible imbalance between outward-directed force from magnetic pressure and the inward force of magnetic tension \citep{priest2014magnetohydrodynamics}. 
A helical current sheet forms and reconnection can occur. The condition for the instability to occur can be expressed in terms of a critical amount of twist $\Phi_{\mathrm{crit.}}$. Different studies have predicted the critical twist in different configurations: $\Phi_{\mathrm{crit.}} = 3.3\,\pi$ for a uniform twisting \citep{hood1979equilibrium}, $\Phi_{\mathrm{crit.}} = 4.8\,\pi$ for a localized twisting profile \citep{mikic1990dynamical}, $\Phi_{\mathrm{crit.}} = 5.15\,\pi$ for localized, variable twisting profile \citep{baty1996electric}.
Energy is released after the magnetic field becomes unstable to ideal MHD modes. At the beginning, a helical kink develops and grows according to the linear theory of instability. Afterwards, the initial helical current sheet progressively fragments, in a turbulent manner, into smaller-scale sheets. During the onset of instability, a burst of kinetic energy occurs. Magnetic dissipation occurs throughout the non-linear phase of the instability. In particular, reconnection events occur in fine-scale structures like current sheets. Dissipation of these sheets is similar to a nanoflare storm. At the end, the magnetic field reaches an energy minimum constrained by conservation of magnetic helicity, as expected in highly conducting plasmas \citep{browning2003solar, browning2008heating}, among other topological constraints \citep[e.g.][]{yeates10topological}.

Since the loop expands in the radial direction, during the relaxation process, an unstable loop can disrupt nearby stable loops \citep{tam2015coronal}, and can trigger an MHD avalanche \citep{2016mhhoodd}. For instance, \cite{2016mhhoodd} demonstrate that an MHD avalanche can occur in a non-potential multi-threaded coronal loop. They show that a single unstable thread can trigger the decay of the entire structure. In particular, each flux tube coalesces with the neighbouring ones and releases discrete heat bursts.
In general, the energy stored by photospheric motions can be released via viscous and Ohmic dissipation during a dynamic relaxation process \citep{reid2018coronal}, and thereafter through a sequence of impulsive, localized and aperiodic heating events under the action of continuous photospheric driving \citep{reid2020coronal}.

Those earlier works conducted exploratory investigations with relatively simplified loop configurations, with no gravitational stratification, and with no variation of $\beta$ with height, and neglecting thermal conduction and optically thin radiation. 
Separately, others have considered the effect of thermal conduction, but in a purely coronal loop \citep{Botha2011}, and again without stratification in density.
In this work, we consider a more realistic scenario of flux tubes interacting within a stratified atmosphere, including chromospheric layers and the thin transition region to the corona, the associated transition from high-$\beta$ to low-$\beta$ regions, and including thermal conduction and optically thin radiation \citep[as in][]{reale20163d}.
The question is whether the avalanche process will be triggered, with what timescales, and how it will develop, as compared with the original, simplified approach.

In section~\ref{sec:model}, we present the equations used in our model and the details for the set-up used in the numerical model. Section~\ref{sec:theory} treats the theoretical models deployed to corroborate the numerical outcomes and to provide a physical interpretation to the results. Section~\ref{sec:results} discusses the results achieved, concerning the initial quasi-static evolution of the coronal loops as they twist under the influence of a slow photospheric motion and the further dynamic relaxation via an MHD avalanche. Finally, in section~\ref{sec:conclusions}, we provide a comprehensive interpretation of the results compared with previous works.
\section{The model} \label{sec:model}

\begin{figure*}[h!]
   \centering
   \includegraphics[height=0.5\hsize]{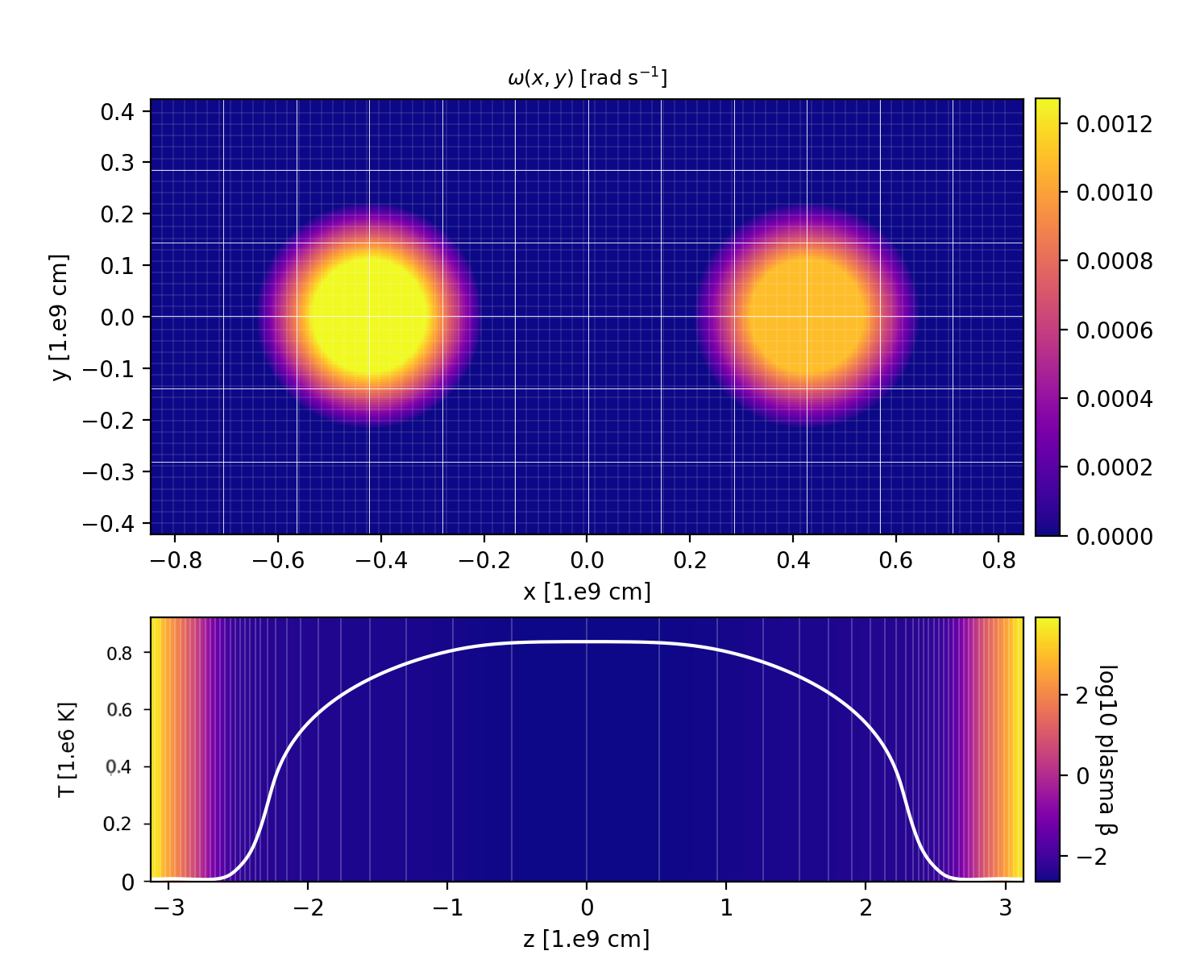}
   \hspace{1 cm}
   \includegraphics[height=0.5\hsize]{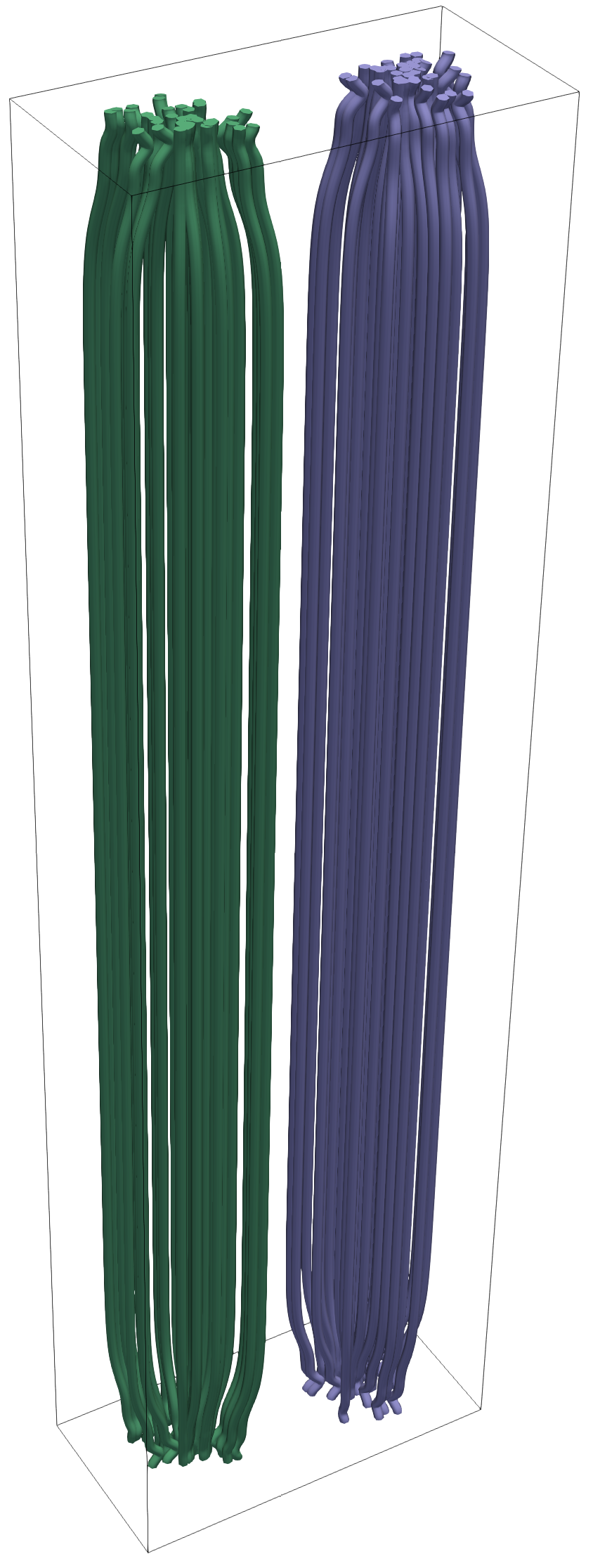}
      \caption{\textbf{Upper left panel}: map of the angular velocity at the bottom of the box. The color scale emphasizes higher angular velocity. The uniform grid is marked. The two rotating regions have the same radius $R_{\mathrm{max}}$. The region on the left has a higher angular velocity ($v_{\mathrm{max., left}} = 1.1 \times v_{\mathrm{max., right}}$). \textbf{Lower left panel}: map of average plasma $\beta$ as function of $\mathrm{z}$ at $t = 0\,\mathrm{s}$,  the solid curve shows the initial temperature along the $\mathrm{z}$ axis. \textbf{Right panel}: 3D rendering of the initial magnetic field configuration in the box around the two flux tubes. The green field lines are twisted faster than the purple ones.}
  \label{Fig:MODEL_init_cond}
\end{figure*}

The numerical experiment is based on a solar atmosphere model that consists of a chromospheric layer and a coronal environment crossed by multiple coronal loop strands.
Each strand is modelled as a straightened magnetic flux tube linked to two chromospheric layers at the opposite ends of a box (Fig.~\ref{Fig:MODEL_init_cond}). The length of each tube is much longer than its cross-sectional radius. Though the loop-aligned gravity is that of a curved, untwisted loop, we neglect other effects of the curvature.
In our scenario, the two (upper and lower) chromospheric layers are the two loop footpoints and are so distant from each other that they can be assumed independent regions. 
 
The evolution of the plasma and magnetic field in the box is described by solving the full time-dependent MHD equations including  gravity (for a curved loop),  thermal conduction (including the effects of heat flux saturation), radiative losses from an optically thin plasma, and an anomalous magnetic diffusivity.
The equations are solved in Eulerian, conservative form:
\begin{align}
\parder{\rho}{t} + \mathbf{\nabla} \cdot (\rho\, \vec v) &= 0,
\label{eq:mass_conservation} \\
\parder{(\rho\, \vec v)}{t} + \mathbf{\nabla} \cdot (\rho \,\vec v\,\vec v) &= - \nabla \cdot (P\,\mathbf{I} + \frac{B^2}{8 \pi}\,\mathbf{I} - \frac{\vec B\, \vec B}{8\ \pi}) \label{eq:momentum} \\
&\qquad + \rho \vec g, \nonumber \\
 \parder{\vec B}{t} - \nabla \times (\vec v \times \vec B) &=  \eta \nabla^2 \vec B,
 \label{induction_eq} \\
\begin{aligned}[t] & \parder{}{t} \left( \frac{B^2}{8 \pi} + \frac{1}{2} \rho v^2 + \rho \epsilon + \rho g h\right) + \\
& \qquad + \nabla \cdot \left[ \frac{c}{4 \pi} \vec E \times \vec B + \frac{1}{2}\rho v^2 \vec v \right. +
\label{eq:energy}
\end{aligned} 
\\
\begin{aligned}
    & \qquad  \left. + \frac{\gamma}{\gamma-1} P\, \vec v + \vec F_c + \rho g h\, \vec v \right] \nonumber 
\end{aligned} &= - \Lambda(T) n_e n_H + H_0, \\ 
    P &= (\gamma -1) \rho \epsilon = \frac{2 k_B}{\mu m_H} \rho T,
    \label{eq: EOS} \\
\vec j &= \frac{c}{4 \pi} \mathbf{\nabla} \times \vec B,
\label{Eq:current_density} \\
    \vec E &= -  \frac{\vec v}{c} \times \vec B + \frac{\vec j}{\sigma},
\label{Eq:electric_field}
\end{align}
where $t$ is the time; 
$\rho$ the mass density; 
$\vec{v}$ the plasma velocity; $P$ the thermal pressure; 
$\vec{B}$ the magnetic field; $\vec E$ the electric field;
$\vec{g}$ 
the gravity acceleration vector for a curved loop;  $\mathbf{I}$ the
identity tensor;
$\epsilon$ the internal energy; 
$\vec{j}$ the induced current density; 
$\eta$ the magnetic
diffusivity;  
$\sigma = \frac{c^2}{4 \pi \eta}$ the electrical conductivity;
$T$ the temperature; 
$\vec{F}_c$ the thermal conductive flux; $\Lambda (T)$  the optically thin radiative losses per unit emission measure; $n_H$ and $n_e$  the hydrogen and electron number density, respectively; $m_H$ the hydrogen mass density; $k_B$  the Boltzmann constant;
$\mu = 1.265$ the mean ionic \textit{weight} \citep[relative to a proton and assuming metal abundance of solar values: $X\,(\mathrm{H}) \simeq 70.7\,\%$, $Y\,(\mathrm{He}) \simeq 27.4\,\%$, $Z\,(\mathrm{Li-U}) \simeq 1.9\,\%$;][]{anders1989abundances}; 
and $H_0 = 4.3 \times 10^{-5}\,\mathrm{erg}\,\mathrm{cm}^{-3}\,\mathrm{s}^{-1}$ a volumetric heating rate which balances the initial energy losses and is used to keep the loop initially in thermal equilibrium.
As shown in Eq. (\ref{eq: EOS}), we use the ideal gas law as an equation of state. 
 
According to the induction equation (Eq. \ref{induction_eq}), the magnetic field soleinodal condition $\mathbf{\nabla} \cdot \vec B = 0$ formally holds at any time $t$, provided the initial conditions are well posed (and numerical errors are not taken into account).
Ampere's law in non-relativistic regime (Eq. \ref{Eq:current_density}) gives the current density in terms of the curl of the magnetic field.
In Eq. (\ref{Eq:current_density}), the displacement current can be neglected provided that the plasma velocity is not relativistic (i.e. provided $v \ll c$).
The electric field $\vec E$ is defined by Ohm's law in equation (\ref{Eq:electric_field}).
From this, the Poynting flux can be decomposed into three terms:
\begin{equation}
\frac{c}{4 \pi} \vec E \times \vec B =  - \frac{1}{4 \pi} \vec B (\vec v \cdot \vec B) + \frac{B^2}{4 \pi} \vec v + \frac{\eta}{c}  \vec j \times \vec B.
\end{equation}
The first term on the right-hand side will be significant for the driving as it determines the energy injected in the domain by the photospheric driver. The second term represents instead the flow of magnetic energy across the boundaries of the domain. Finally, the third term is related to Ohmic dissipation and field line diffusion at the boundaries of the domain.

\subsection{Thermal conduction, radiative losses and heating}
\label{sec:thermal}
The thermal conductive flux is defined in the equations below. 
\begin{align}
\vec{F}_c &= \frac{F_{\mathrm{sat.}}}{F_{\mathrm{sat.}} + |F_{\mathrm{class.}}|} \vec{F}_{\mathrm{class.}}, \\
\vec{F}_{\mathrm{class.}} &= - k_{\parallel} \hat{\vec{b}} ( \hat{\vec{b}} \cdot \mathbf{\nabla} T) - k_{\perp} \left[ \mathbf{\nabla} T - \hat{\vec{b}} ( \hat{\vec{b}} \cdot \mathbf{\nabla} T) \right], \\
F_{\mathrm{sat.}} &= 5 \phi \rho c^3_{\mathrm{iso.}}.
\end{align}
where, the subscripts $\parallel$ and $\perp$ denote the components parallel and perpendicular to the magnetic field.
The thermal conduction coefficients along and across the field are $k_{\parallel} = K_{\parallel} T^{\frac{5}{2}}$ and
$k_{\perp} = K_{\perp} \rho^2 / (B^2T^{\frac{1}{2}})$, where
$K_{\parallel} = 9.2 \times 10^{-7}$ and $K_{\perp} = 5.4 \times 10^{-16}$ (cgs units); $c_{\mathrm{iso.}}$ is the isothermal sound speed; $\phi = 0.9$ is a dimensionless free parameter; $\hat{\vec{b}} = \vec B / B$ a unit vector in the direction of the magnetic field; and $F_{\mathrm{sat.}}$ is the maximum flux magnitude in the direction of $F_c$.

The optically thin radiative losses per unit emission measure are derived from the CHIANTI v.~7.0 database \citep{landi2013prominence}, assuming coronal element abundances \citep{widing1992element}.
 
Across the transition region the number density and the temperature change by three orders of magnitude in less than $100\,\mathrm{km}$. Resolving such rapid variation and steep gradients would ordinarily require an extremely high spatial resolution and lead to unfeasible computational times \citep{bradshaw2013influence}. 
The Linker--Lionello--Mikić method \citep{linker2001magnetohydrodynamic,lionello2009multispectral,mikic2013importance} allows us artificially to broaden the transition region without significantly changing the properties of the loop in the corona, obviating that challenge.
In particular, following the Linker-Lionello--Mikić approach, we modified the temperature dependence of the parallel thermal conductivity and radiative emissivity below a  temperature threshold, $T_c = 2.5 \times 10^5\,\mathrm{K}$: 
\begin{align}
    \tilde{k}_{\parallel}(T) &= \begin{cases} k_{\parallel}(T)\; , & T > T_c\; , \\ k_{\parallel} (T_c)\; , & T < T_c\; , \end{cases} \\
   \tilde{\Lambda}_{\parallel} (T < T_c) &= \begin{cases} \Lambda_{\parallel} (T)\; , & T > T_c\; , \\ \Lambda_{\parallel} (T) \left( \frac{T}{T_c} \right)^{5/2}\; , & T < T_c\; . \end{cases}
\end{align}
According to Rosner-Tucker-Vaiana scaling laws \citep{rosner1978dynamics}, the volumetric heating rate $H_0$ is sufficient to keep the  corona static with an apex temperature of about $8 \times 10^{5}\,\mathrm{K}$ and half length $L = 2.5 \times 10^9\,\mathrm{cm}$. This provides a background atmosphere that is adopted as initial conditions, according to the hydrostatic loop model by \cite{serio1981closed} and \cite{guarrasi2014mhd}. 
That heating rate is not similarly scaled for temperatures below the cut-off temperature \citep[cf.][]{johnston2020modelling}

\subsection{Gravity in a curved loop}

We assume that the flux tube is circularly curved only in the corona, and that it is straight in the chromosphere. Thus we consider the gravity of a curved loop in the corona:
\begin{equation}
    g(z) \,\hat{\mathbf{z}} = g_{\odot} \sin{\left(\pi \frac{z}{L}\right)} \,\hat{\mathbf{z}},
\end{equation}
where $g_{\odot} = \frac{G M_{\odot}}{R_{\odot}^2}$ is constant; $G$ is the gravitational constant; $M_{\odot}$ is the solar mass; $R_{\odot}$ is the solar radius. 
Note that gravitational acceleration decreases and becomes zero at the loop apex ($z = 0$), to account for the loop curvature. Below the corona, gravity is uniform and vertical.

\subsection{The loop setup}

Our computational 3D box contains two flux tubes of length $5 \times 10^9\,\mathrm{cm}$ and initial temperature of approximately $10^6\,\mathrm{K}$ (see lower-left panel of figure \ref{Fig:MODEL_init_cond}). Their footpoints are anchored to two thick, isothermal chromospheric layers, at the top and bottom of the box. As the plasma $\beta$ decreases farther from the boundaries, the magnetic field expands (see Fig. \ref{Fig:MODEL_init_cond}). The initial atmosphere is the result of a preliminary simulation in which we let a box with a vertical magnetic field relax to an equilibrium condition until the maximum velocity has reached a value below $10\,\mathrm{km}\,\mathrm{s}^{-1}$, as described in \cite{guarrasi2014mhd}.

The computational box is a 3D Cartesian grid,  $-x_M < x < x_M$, $-y_M < y < y_M$, and $-z_M < z < z_M$, where $x_M = 2 y_M = 8 \times 10^8\,\mathrm{cm}$, $y_M = 4 \times 10^8\,\mathrm{cm}$, and $z_M = 3.1 \times 10^9\,\mathrm{cm}$, with a staggered grid.
We adopted the piecewise uniform and stretched grid sketched in Fig.~\ref{Fig:MODEL_init_cond}.
In particular, in the corona, we considered a non-uniform grid whose resolution degrades with height. Instead, to describe the transition region at sufficiently high resolution, the cell size there ($|z| \approx 2.4 \times 10^9\,\mathrm{cm}$) decreases to $\Delta r \sim \Delta z \sim 3 \times 10^6\,\mathrm{cm}$ and remains constant across the chromosphere.  The boundary conditions are periodic at $x = \pm x_M$ and $y = \pm y_M$; and reflective but with reversed sign for the tangential component of the magnetic field (i.e. equatorial symmetric boundary conditions) at $z = \pm z_M$.

We also performed a second numerical experiment, extending the domain in the x-direction ($x_M = 1.5 \times 10^9\,\mathrm{cm}$) and including a third flux tube. The results of this simulation are discussed in section~\ref{sec:conclusions}.
 
\subsection{The plasma resistivity}

We consider an anomalous plasma resistivity that is switched on only in the corona and transition region (i.e. above $T_{\mathrm{cr.}} = 10^4\,\mathrm{K}$) where the magnitude of the current density exceeds a critical value, as in the following equation \citep[e.g.][]{hood2009coronal}:
\begin{equation}
    \eta =
    \begin{cases}
    \eta_0 & |J| \ge J_{\mathrm{cr.}}  \text{ and } T \ge T_{\mathrm{cr.}} \\
    0  & |J| < J_{\mathrm{cr.}} \text{ or } T < T_{\mathrm{cr.}}
    \end{cases},
\end{equation}
where we assume $\eta_0 = 10^{14}\,\mathrm{cm}^{-2}\,\mathrm{s}^{-1}$ and $J_{\mathrm{cr.}} = 250\,\mathrm{Fr}\,\mathrm{cm}^{-3}\,\mathrm{s}^{-1}$. The current threshold has been chosen so as to avoid Ohmic heating before the onset of the instability and to permit the ideal build-up to the instability. With this assumption, the minimum heating rate above threshold is $H = \eta_0 (4 \pi |J_{\mathrm{cr.}}|/c)^2 \approx 0.3\,\mathrm{erg}\,\mathrm{cm}^{-3}\,\mathrm{s}^{-1}$.  
Below the critical current, a minimum numerical resistivity is inevitably present, but it does not contribute any heating during the simulation.

\subsection{Loop twisting}

We tested the evolution of a coronal loop under the effects of a footpoint rotation. In particular, both strands are driven by coherent photospheric rotations which switch sign from one footpoint to the other. Rotation at the threads' footpoints induces a twisting of the magnetic field lines. As flux tube torsion increases, the current density is amplified. Once the conditions for a kink instability are reached, a strong current sheet forms and the critical current is exceeded, triggering magnetic diffusion and heating via Ohmic dissipation. The angular velocity $\omega(r)$ is that of a rigid body around the central axis, i.e., the angular speed is constant in an inner circle and then decreases linearly in an outer annulus \citep{reale20163d}: 
\begin{equation}
    \omega = \omega_0 \times
    \begin{cases}
    1 & r < r_{\mathrm{max.}}   \\
    (2 r_{\mathrm{max.}} - r)/r_{\mathrm{max.}} & r_{\mathrm{max.}} < r < 2 r_{\mathrm{max.}}    \\
    0 & r > 2 r_{\mathrm{max.}}
    \end{cases},
    \label{eq:angular_vel}
\end{equation}
where $\omega_0 = v_{\mathrm{max.}}/r_{\mathrm{max.}}$, $v_{\mathrm{max.}}$ is the maximum tangential velocity ($v_{\phi} = \omega r$) and $r_{\mathrm{max.}}$ is the characteristic radius of the rotation. 
In this specific case, the central loop is driven at a speed which is 10\% higher than the lateral ones and is equal to $1.1\,\mathrm{km}\,\mathrm{s}^{-1}$. The maximum velocity achieved by twisting is always smaller than the minimum Alfvén velocity $v_\mathrm{A} = B/\sqrt{4 \pi \rho}$ in the domain. Moreover, the characteristic velocity ($\omega_0 \, r_{\mathrm{max.}}$) is high enough to avoid field line slippage at the photospheric boundaries caused by numerical diffusion. The choice of a mirror-symmetric photospheric driver does not make the further system evolution to lack of generality: the relatively high Alfvén velocities lead coronal loops to maintain a very high degree of symmetry even when they are subjected to asymmetric photospheric motions for a long time \citep{sym15030627}.
The $r_{\mathrm{max.}}$ parameter is set to $1200\,\mathrm{km}$ for both loops (see top-left panel of figure \ref{Fig:MODEL_init_cond}).
For many recent simulations of coronal loops subject to photospheric driving, a significant challenge has been attaining realistic driving speeds, with this need to drive quickly enough to prevent field line slippage requiring modelled velocities much faster than those observed.
However, with velocities of the order of $1\,\mathrm{km}\,\mathrm{s}^{-1}$, we approach typical photospheric velocity patterns \citep{gizon2005local,rieutord2010Sun}, benefiting from growing computational resources.

\subsection{Numerical computation}

The calculations are performed using the PLUTO code \citep{mignone2007pluto, mignone2012conservative}, a modular, Godunov-type code for astrophysical plasmas. The code provides an algorithmic, modular multiphysics environment, particularly oriented toward the treatment of astrophysical flows in the presence of discontinuities as in the case treated here. 
Numerical integration of the conservation laws (Eqs. \ref{eq:mass_conservation}--\ref{eq:energy}) is achieved through high-resolution shock-capturing (HRSC) schemes using the finite volume (FV) formalism where volume averages evolve in time \citep{mignone2007pluto}.
The code is designed to make efficient use of massive parallel computers using the message passing interface (MPI) library for interprocessor communications. The MHD equations are solved using the MHD module available in PLUTO, configured to compute intercell fluxes with the Harten-Lax-Van Leer approximate Riemann solver \citep{roe1986characteristic}, while second order accuracy in time is achieved using a Runge-Kutta scheme. A Van Leer limiter \citep{sweby1985high} for the primitive variables is used. The evolution of the magnetic field is carried out adopting the constrained transport approach \citep{balsara1999staggered} that maintains the solenoidal condition ($\mathbf{\nabla} \cdot \vec{B} = 0$) at machine accuracy.
The PLUTO code includes optically thin radiative losses in a fractional step formalism, which preserves the second-order time accuracy, since the advection and source steps are at least second-order accurate; the radiative losses, $\Lambda$, values are computed at the temperature of interest using a table look-up/interpolation method. Thermal conduction is treated separately from advection terms through operator splitting. In particular, we adopted the super-time-stepping technique \citep{alexiades1996super}, which has proved to be very effective at speeding up explicit time-stepping schemes for parabolic problems. 
This approach is crucial when high values of plasma temperature are reached (e.g., during flares) since explicit schemes are subject to a rather restrictive stability condition (namely, $\Delta t \le (\Delta x)^2/(2 \alpha)$, where $\alpha$ is the maximum diffusion coefficient). This means the thermal conduction timescale, $\tau_{\mathrm{cond.}}$, is shorter than the dynamical one, $\tau_{\mathrm{dyn.}}$ \citep{orlando2005crushing,orlando2008importance}.

\section{Basic theory} \label{sec:theory}
\subsection{Twisting with expanding magnetic tube}

Simple analytical models can predict the initial, steady state evolution of the system provided that certain assumptions be satisfied \citep{hood1979equilibrium, browning1989shape}. Each loop is modelled as a cylindrically symmetric magnetic structure not interacting with the neighbouring ones. 
The initial magnetic field is not uniform: magnetic field lines expand from the photospheric boundaries (where $B \approx 300\,\mathrm{G}$) to the upper corona (where $B \approx 10\,\mathrm{G}$). As the field lines expand, the magnetic field decreases by an order of magnitude because of conservation of magnetic flux.
The expansion of the field corresponds to a height that is roughly equal to the distance between the chromospheric sources, so it involves a small fraction of a coronal loop length and thus the loop is of mostly uniform width in the corona.
Field line tapering is strong in the chromosphere, where changes in the plasma beta are steeper, but weaker in the corona. In typical coronal conditions, such as high ($T \approx 1\,\mathrm{MK}$) and uniform temperature, the pressure scale height is large compared with the loop length. Therefore, density and pressure can be assumed to be uniform and constant in the corona. Averaged values can be constrained from the RTV scaling laws \citep{rosner1978dynamics} once the total length of the loop $2 L$ and the uniform heating rate $H_0$ are given. At the boundaries, the photospheric driver twists the magnetic field, causing the azimuthal component to increase. Perfect line-tying to the photospheric boundaries is assumed and no field line slippage is taken into account.  The driver is much slower than the Alfvén velocity so that the magnetic torsion can be assumed to be instantaneously transmitted along the whole tube.
 
Under these assumptions, it is possible to express the magnetic field of a single thread in cylindrical coordinates in terms of the flux function $\psi$ as a generalized parameter:
\begin{equation}
    \vec B = \frac{1}{R} \left( - \parder{\psi}{z}, G(\psi), \parder{\psi}{R} \right),
    \label{eq:mag_filed_cyl}
\end{equation}
where $\psi(R, z) = \int_0^R B_z(r^\prime , z) r^\prime\,\mathrm{d}r^\prime$. 
In equation (\ref{eq:mag_filed_cyl}), the azimuthal magnetic field component is given by a function, $G$, of the flux function, $\psi$. $G$ is related to the radial profile of the twisting velocity. In the following, $R$ will denote the radius in the corona and $r$ the radius in the photosphere. 
Under typical coronal conditions (i.e. low plasma beta), the force-free condition:
\begin{equation}
    (\nabla \times \vec B) \times \vec B = 0\; ,
\end{equation}
holds.
Using Eq. (\ref{eq:mag_filed_cyl}), the force-free equation is satisfied by the Grad-Shafranov equation \citep{GradRubin1958,Shafranov1958}:
\begin{equation}
    \frac{\partial^2 \psi}{\partial R^2} - \frac{1}{R} \parder{\psi}{R} + \frac{\partial^2 \psi}{\partial z^2} + F(\psi) = 0,
\end{equation}
where $F = G \frac{\mathrm{d}G}{\mathrm{d}\psi}$ is a function of $\psi$. The third term on the left-hand side can be neglected in the corona, under the assumption of negligible field line curvature, $\parder{\psi}{z} \approx 0$ \citep{browning1989shape,lothian1989twisted}.
 As a first assumption, we consider the azimuthal velocity as linear in $z$, since the twisting is equal in magnitude and opposite in direction at the two ends:
 \begin{equation}
     v_{\phi} = \omega(r) r
     \frac{z}{l}
     \label{eq:velocity_tube},
 \end{equation}
 where $l$ is a length scale that can be assumed equal to $L$, the half-length of the loop, in order that this equation match the angular speed imposed on the boundaries.
 From the linearization of the ideal induction equation (Eq. \ref{induction_eq}), with $\eta = 0$, the azimuthal component of the magnetic field in the photosphere is easily linked to the given twisting angular velocity $\omega (r)$ (see Eq. (\ref{eq:angular_vel})):
\begin{equation}
    \parder{B_{\phi, \text{phs}}}{t} = \parder{\left(B_z (r) v_{\phi}\right)}{z} = \frac{B_z(r) \omega (r) r}{l} t = r G.
    \label{eq:induction_phi}
\end{equation} 
The vertical component at the photosphere, $B_{z, \text{phs}} (r)$, is given by the superposition of a background magnetic field $B_{\mathrm{bk}}$ and a Gaussian function with amplitude $B_0$ and a characteristic width $\varsigma$, i.e.:
\begin{equation}
    B_{z, \text{phs}}(r) = B_{\mathrm{bk}} + B_0 e^{-\frac{r^2}{\varsigma^2}} 
\end{equation}
and so:
\begin{equation}
\psi(r, \pm L) = \frac{1}{2}B_{\mathrm{bk}}r^2 - \frac{B_0 \sigma^2}{2}\left (e^{-r^2/\varsigma^2} - 1\right )\; .
\end{equation}

The magnetostatic equilibrium of a coronal loop in response to slow twisting of the photospheric footpoints can be investigated in the corona by solving the Grad-Shafranov equation:
\begin{align}
    B_{\phi, \text{coro.}} (R) &= \frac{G}{R} \\
    B_{z, \text{coro.}} (R) &= \frac{1}{R} \parder{\psi}{R}\;.  
\end{align}
In this way we account for the flux tube expansion across the chromosphere just by assuming magnetic flux conservation throughout the loop volume and force free condition in the corona. 
In particular, the volume-integrated magnetic energy of a single thread in the corona is given by:
\begin{equation}
    E_{\mathrm{mag.}} = 2 \pi \cdot 2L \cdot \int_0^{y_M}  r\frac{B_{z \text{coro.}}^2 + B_{\phi, \text{coro.}}^2}{8 \pi}\,\mathrm{d}r.
\end{equation}
The volume-integrated kinetic energy can be roughly assessed by assuming the coronal loop reaches a steady state where the plasma is moved only by the magnetic field torsion:
\begin{equation}
    E_{\mathrm{kin.}} = 2 \pi \int_L^{-L}\,\mathrm{d}z \cdot \int_0^{y_M} r \frac{1}{2} \langle \rho \rangle \Omega (r)^2 \left(\frac{z}{L}\right)^2 r^2\,\mathrm{d}r,
    \label{eq:kin_energy}
\end{equation}
where $\Omega$(r) is the angular velocity of the loop in the corona (the relation $\Omega (\psi) = \omega(\psi)$ holds) and $\langle \rho \rangle$ is the averaged coronal density.

In a cylindrically symmetric flux tube, the angular velocity produces the axial current density and the azimuthal magnetic field (see Eq. (\ref{eq:angular_vel})):
\begin{equation}
J_z = \frac{1}{r} \parder{}{r} \left(r B_{\phi} (r) \right) = \frac{c}{4 \pi} \frac{2 \omega_0 B_z t}{L} \times
    \begin{cases}
        1 & r < r_{\mathrm{max.}}   \\
        \frac{4 r_{\mathrm{max.}} - 3 r}{2 r_{\mathrm{max.}}} & r_{\mathrm{max.}} < r < 2 r_{\mathrm{max.}}    \\
        0 & r > 2 r_{\mathrm{max.}}
    \end{cases},
\end{equation}
according to Eq. (\ref{Eq:current_density}), (\ref{eq:angular_vel}), (\ref{eq:velocity_tube}) and (\ref{eq:induction_phi}).
A rough but effective estimate of the maximum current density over the time is retrieved by evaluating the current density at the loop axis:
\begin{equation}
    J_{\mathrm{max.}} = J_z(r=0,t) = \frac{c}{4 \pi} \frac{2 B_z(0) \omega (0) }{L} \times t.
    \label{eq:max_current}
\end{equation}
As soon as the azimuthal component of the magnetic field increases linearly with time, magnetic energy should grow quadratically and the current density linearly.

\subsection{Energy equations}

\begin{figure}[t]
   \centering
  \includegraphics[width=\hsize]{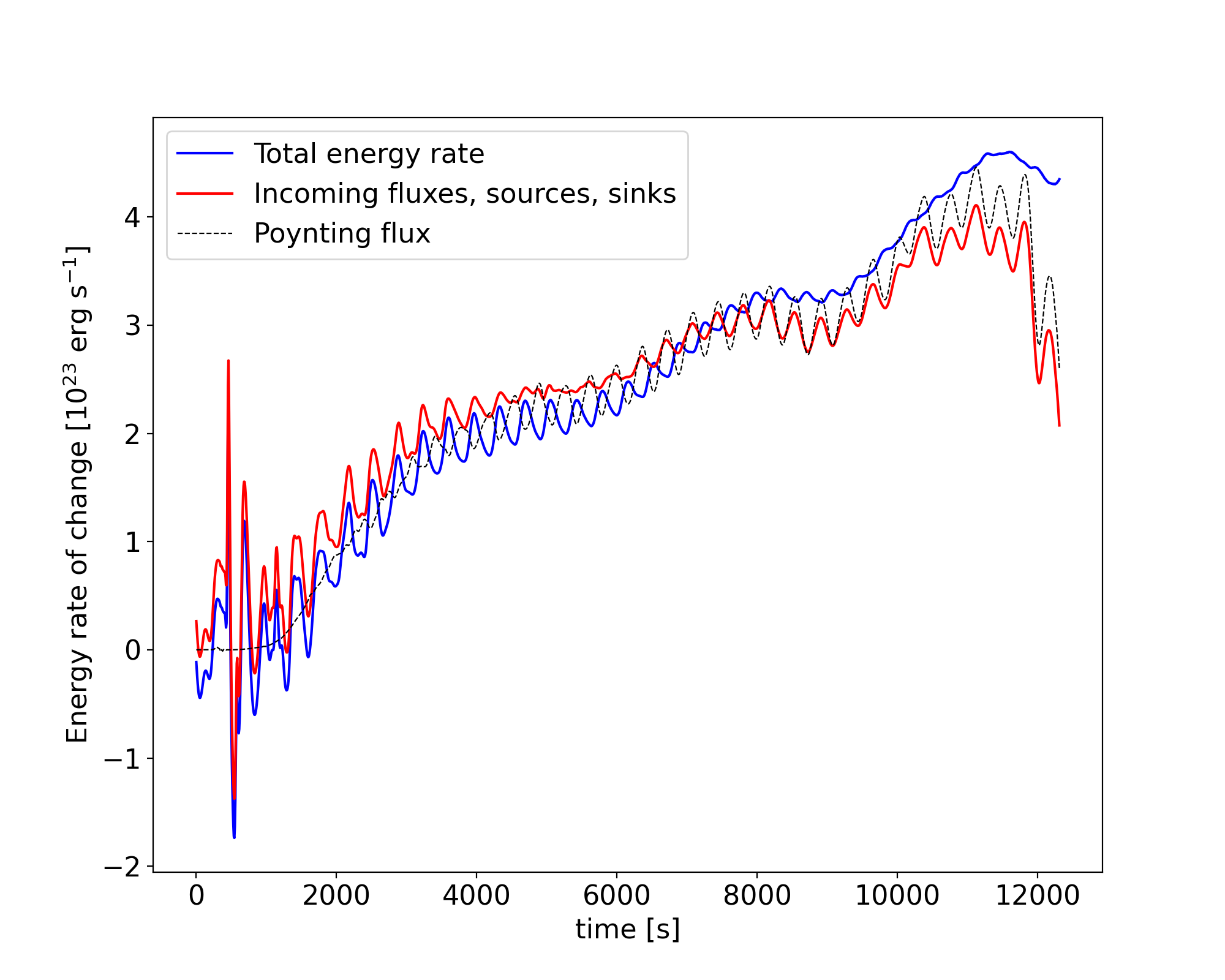}
  \caption{\textbf{Solid, blue curve}: rate of change of the total energy, given by the sum of internal, kinetic, magnetic, and gravitational energies of the system, plotted as functions of time, before the onset of the instability.
  \textbf{Solid, red curve}: sum of the total fluxes, energy sources and sinks as a function of time, before the onset of the instability. Closeness of the blue and red curves demonstrates approximate energy conservation in the domain. \textbf{Dashed, black curve}: the dominant flux is the Poynting flux, which adds to magnetic energy.}
  \label{Fig:RESULTS_rate_of_change}
\end{figure}
\begin{figure}[t]
   \centering
  \includegraphics[width=\hsize]{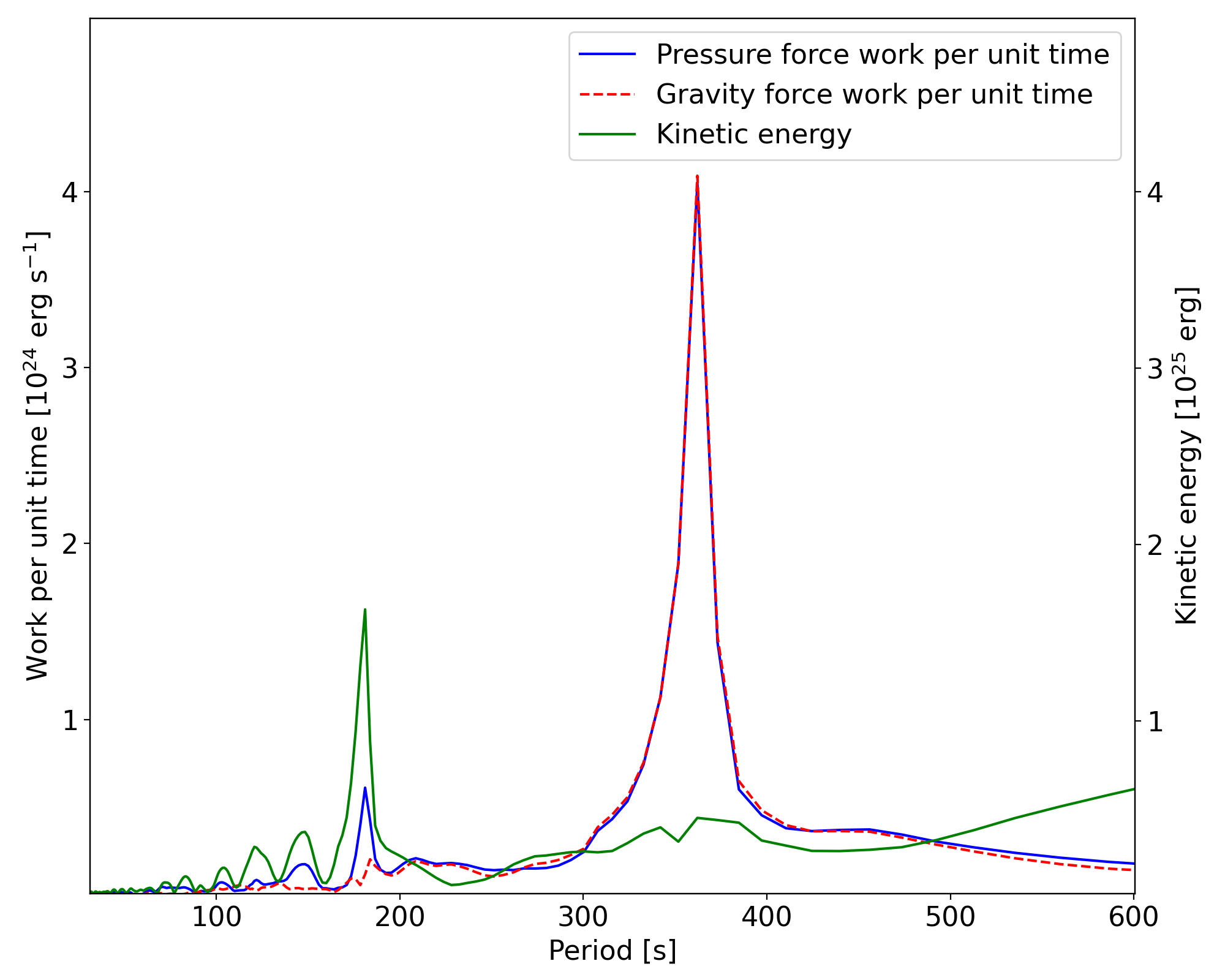}
  \caption{\textbf{Solid, blue curve}: Fourier transform of the work done by pressure forces per unit time, before the onset of the instability.
  \textbf{Solid, red curve}:  Fourier transform of the work done by gravity force per unit time, before the onset of the instability. Both curves show a peak around $T \approx  365\,\mathrm{s}$
 (identified by eye).
 \textbf{Solid, green curve}: Fourier transform of the kinetic energy before the onset of the instability. It shows a peak around $180\,\mathrm{s}$.}
  \label{Fig:RESULTS_BV_frequency}
\end{figure}

\begin{figure}[t]
\begin{center}
  \includegraphics[width=\hsize]{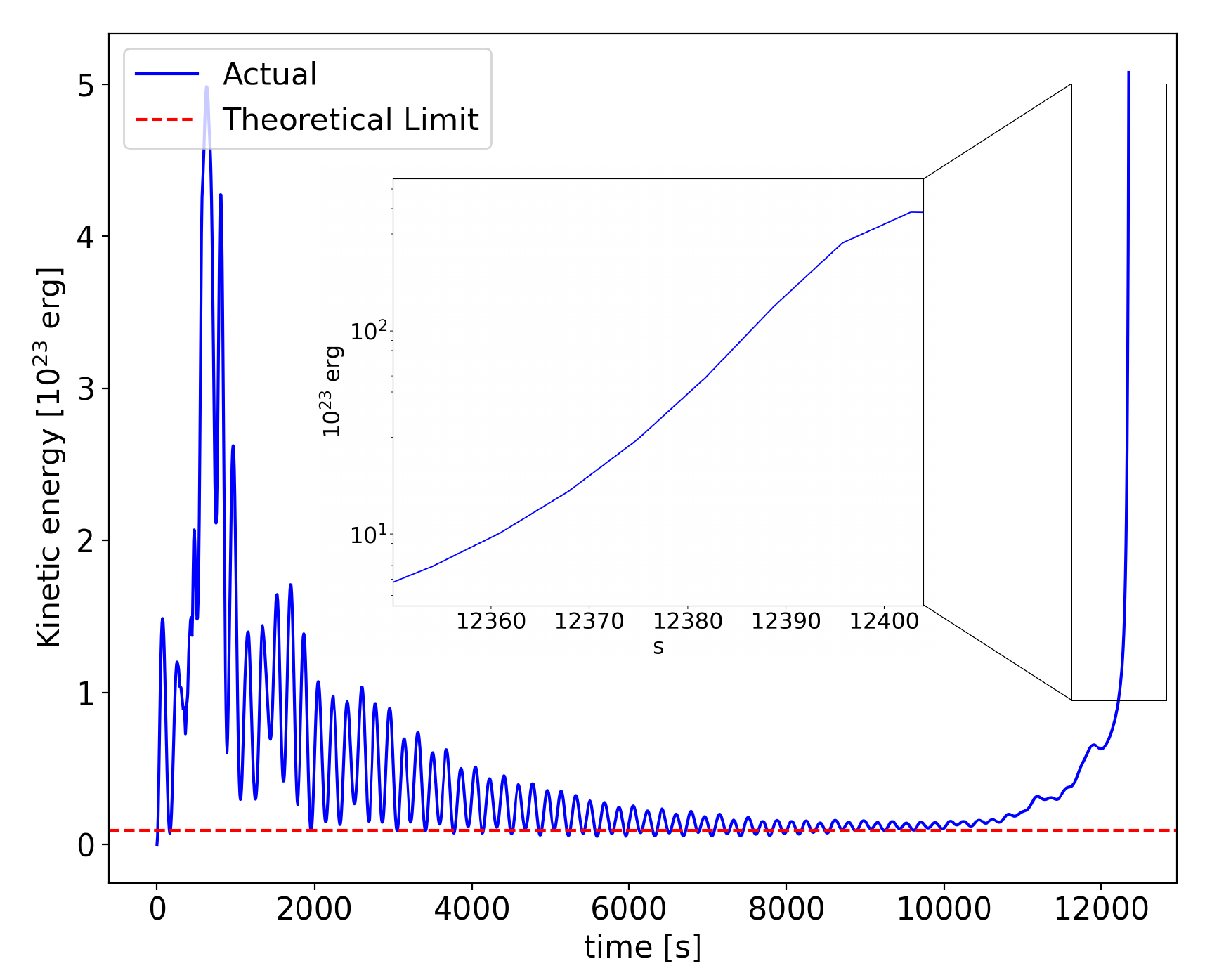}
  \caption{\textbf{Solid, blue curve}:  Total kinetic energy plotted as functions of time, before the onset of the instability. The last exponential rise shows the time at which the first thread is disrupted. \textbf{Dashed, red curve}: Theoretical estimation of the steady state based on the model described in Eq. (\ref{eq:kin_energy}). The system (solid, blue curve) is expected to converge in the linear phase of evolution to that trend as waves are progressively damped.}
  \label{Fig:RESULTS_kinetic_energy}
 \end{center}
\end{figure}

The evolution of total energy is determined by the several energy processes accounted by the numerical simulation. The temporal evolution of the four energy terms (i.e. magnetic, kinetic, internal, and gravitational energy) is driven by the energy sources and sinks (background heating and radiative losses, respectively) and several energy fluxes at the boundaries of the domain (such as thermal conduction, Poynting flux, enthalpy flux, and kinetic and gravitational energy fluxes). In addition, energy transfer terms may link two different forms of energy. This is the case of Ohmic heating, converting magnetic energy into heat, or work done by the Lorentz force, pressure gradient and gravity per unit time, converting kinetic energy to magnetic, thermal, and gravitational energy, respectively.
The equations governing the evolution of magnetic, kinetic, internal and gravitational energy, respectively, are as follows: 

\begin{align}
    \begin{aligned}[t] & \parder{}{t} \frac{B^2}{8 \pi} + \nabla \cdot \left[ - \frac{1}{4 \pi} \vec B \left( \vec v \cdot \vec B \right) \right.
    \end{aligned} \label{eq:magnetic_energy}
    \\
    \begin{aligned}[t]
    &\qquad + \left. \frac{B^2}{4 \pi} \vec v + \frac{\eta}{c} \vec j \times \vec B\right] \end{aligned} &= - \frac{j^2}{\sigma} - \frac{\vec v}{c} \cdot \left(\vec j \times \vec B \right), \nonumber \\
\parder{}{t} \left(\frac{1}{2} \rho v^2 \right) + \nabla \cdot \left(\frac{1}{2} \rho v^2 \,\vec v \right)
&= - \vec v \cdot \nabla P + \frac{\vec v}{c} \left( \vec j \times \vec B \right)\label{eq:kinetic_energy} \\
&\qquad + \rho\, \vec v \cdot \vec g, 
     \nonumber
\end{align} 

\begin{align}
\parder{(\rho \epsilon)}{t} + \nabla \cdot \left[ \frac{\gamma}{\gamma-1} P \,\vec v  + \nabla \cdot \vec F_c  \right] &= \vec v \cdot \nabla P  - \Lambda(T) n_e n_H \label{eq:internal_energy} \\
&\qquad + \frac{j^2}{\sigma} + H_0, \nonumber
     \\
       \parder{(\rho g h)}{t} +  \nabla \cdot \left(\rho g h\, \vec v \right)  &=  - \rho\, \vec v \cdot \vec g .\label{eq:gravitational_energy}
\end{align}

The sum of the four equations gives the energy equation (\ref{eq:energy}) discussed in section~\ref{sec:model}. Terms on the left-hand sides include rates of change in energy (the derivatives with respect to time) and energy fluxes (i.e. surface terms, which here appear as divergences). Energy transfer terms, sources and sinks are on the right-hand sides.

\section{Results} \label{sec:results}

\begin{figure}[t]
\begin{center}
  \includegraphics[width=\hsize]{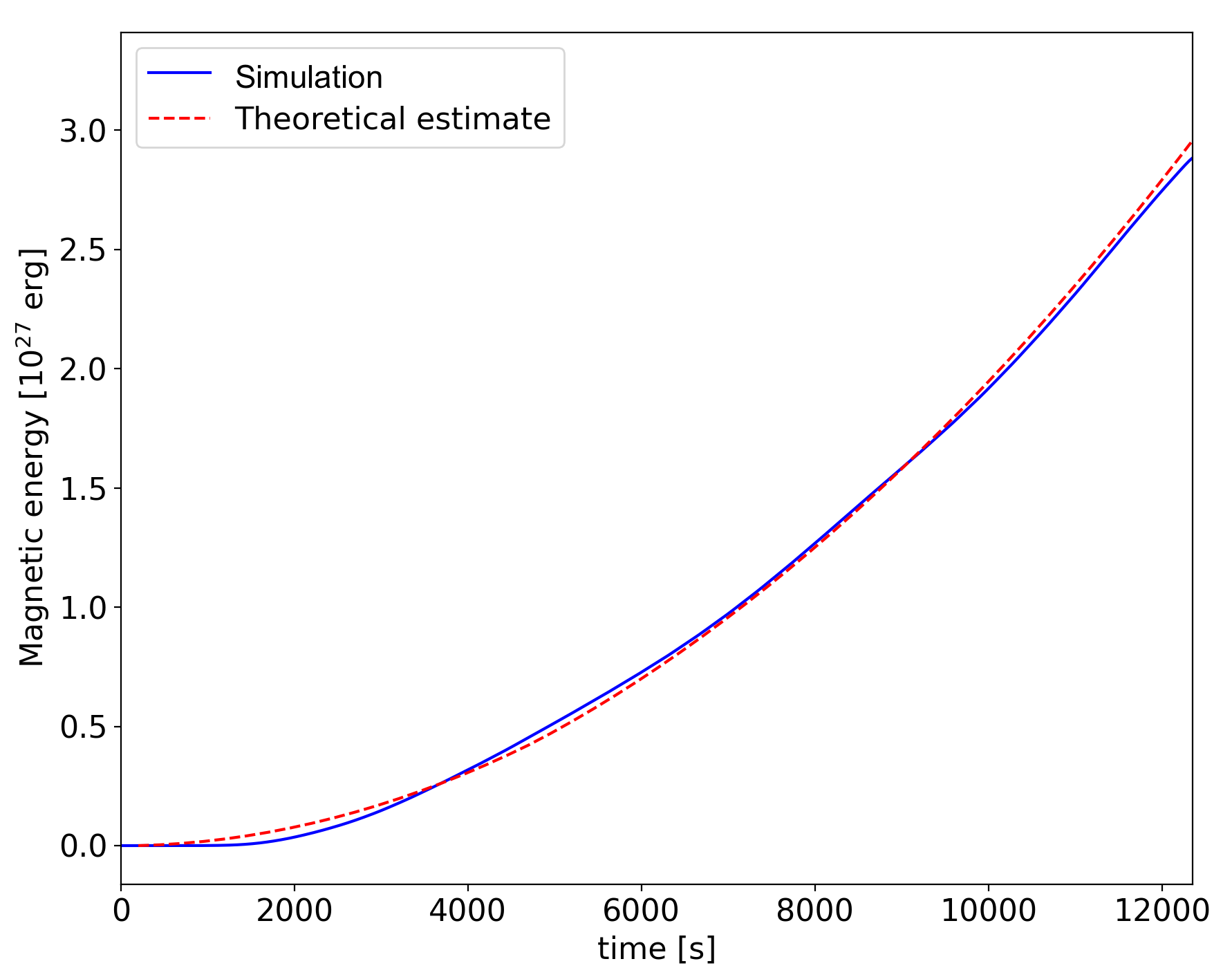}
  \caption{\textbf{Solid, blue curve}:  Total magnetic energy plotted as a function of time, before the onset of the instability. \textbf{Dashed, red curve}:  Theoretical estimate based on the model described in section~\ref{sec:theory}, which grows through the energy input by photospheric driving.}
  \label{Fig:RESULTS_magnetic_energy}
 \end{center}
\end{figure}

\begin{figure}[t]
\centering
  \includegraphics[width=\hsize]{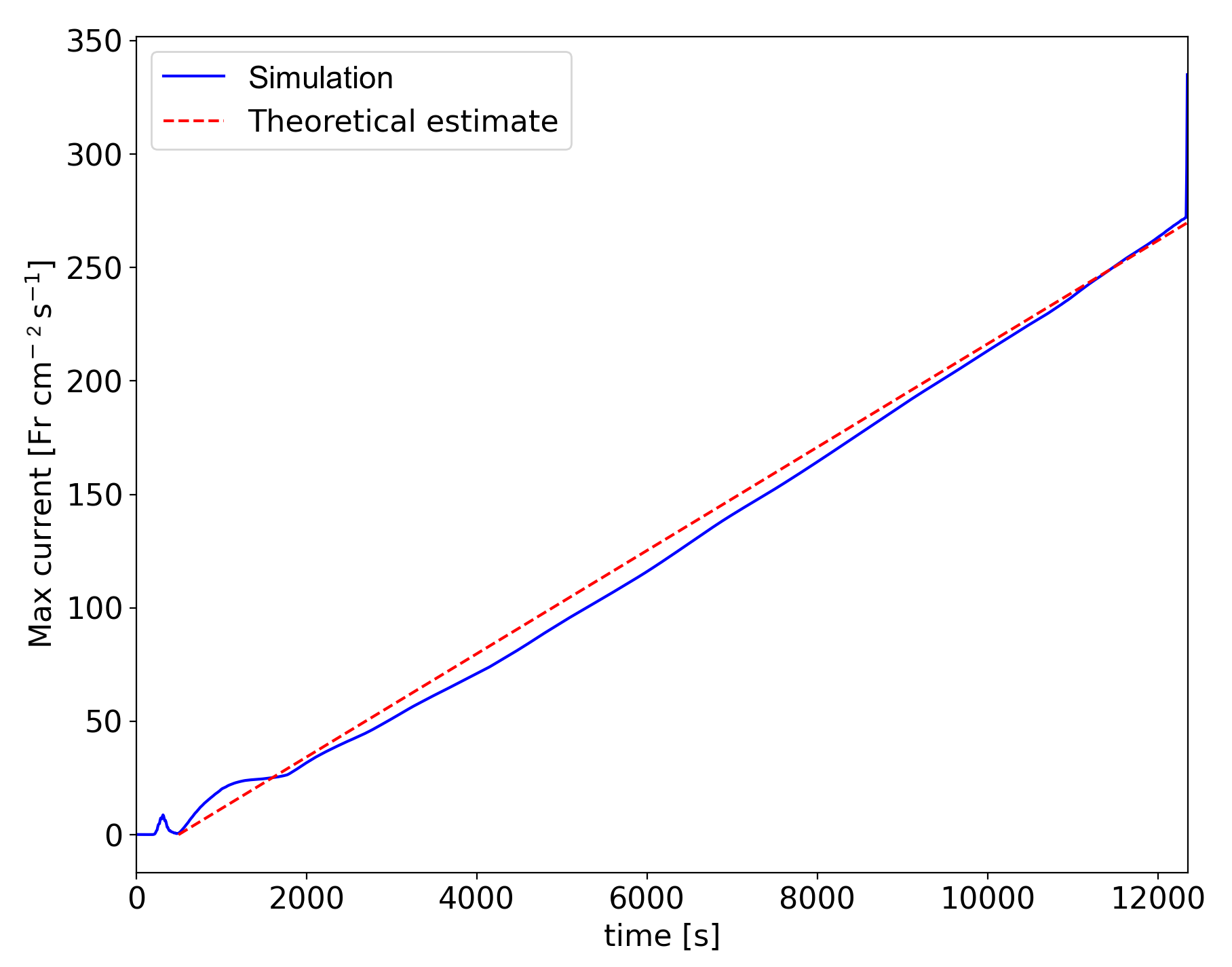}
  \caption{\textbf{Solid, blue curve}:  Maximum current intensity as a function of time, before the onset of the instability. \textbf{Dashed, red curve}:  Theoretical estimate based on the model described in section~\ref{sec:theory}.}
  \label{Fig:RESULTS_max_current}
\end{figure}

\begin{figure}[t]
\centering
  \includegraphics[width=\hsize]{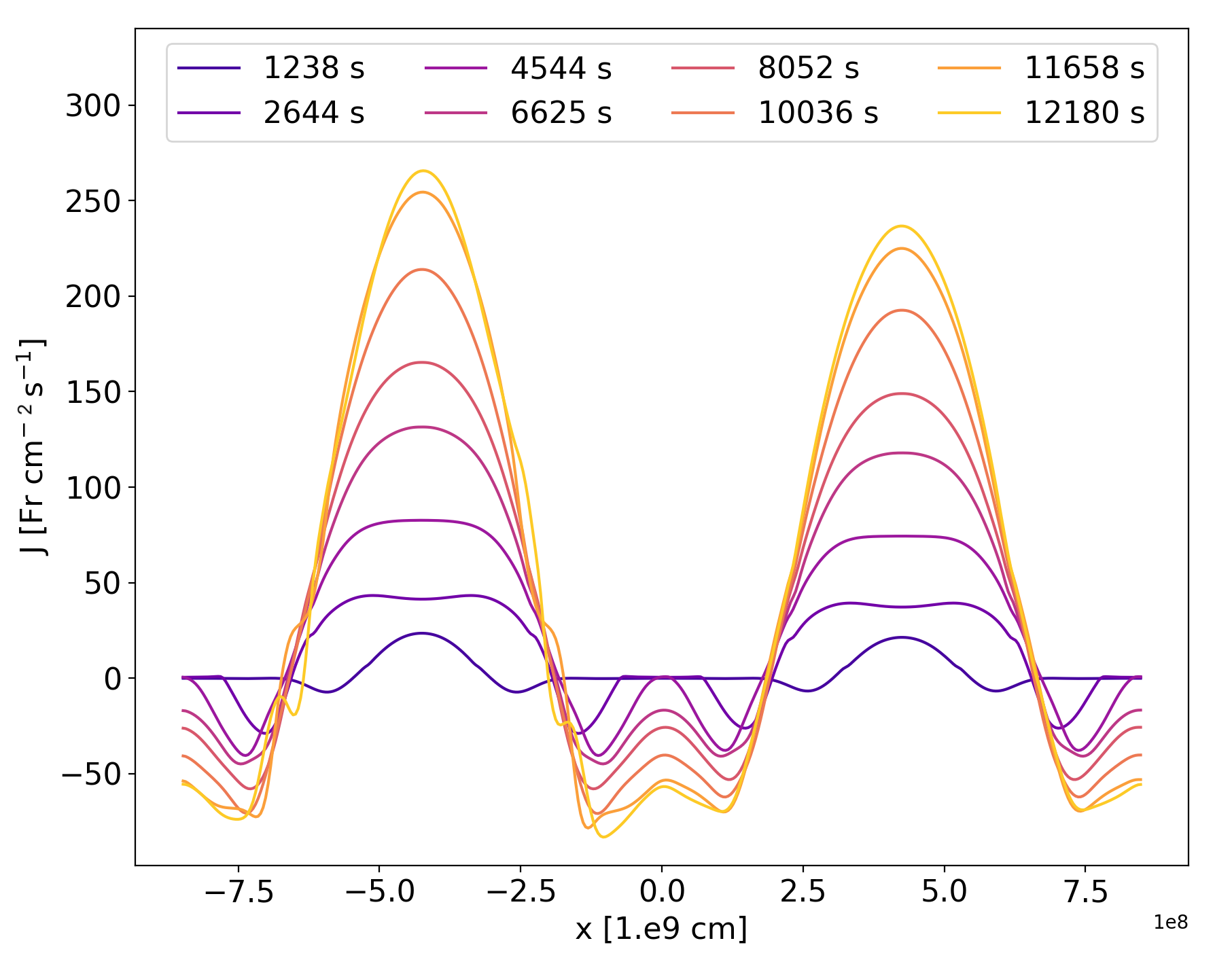}
  \caption{Profiles of the apex current intensity along the x-axis ($y=0$) at different times.}
  \label{Fig:RESULTS_max_current_2}
\end{figure}

\begin{figure*}[!]
\centering
  \includegraphics[width=0.3\hsize]{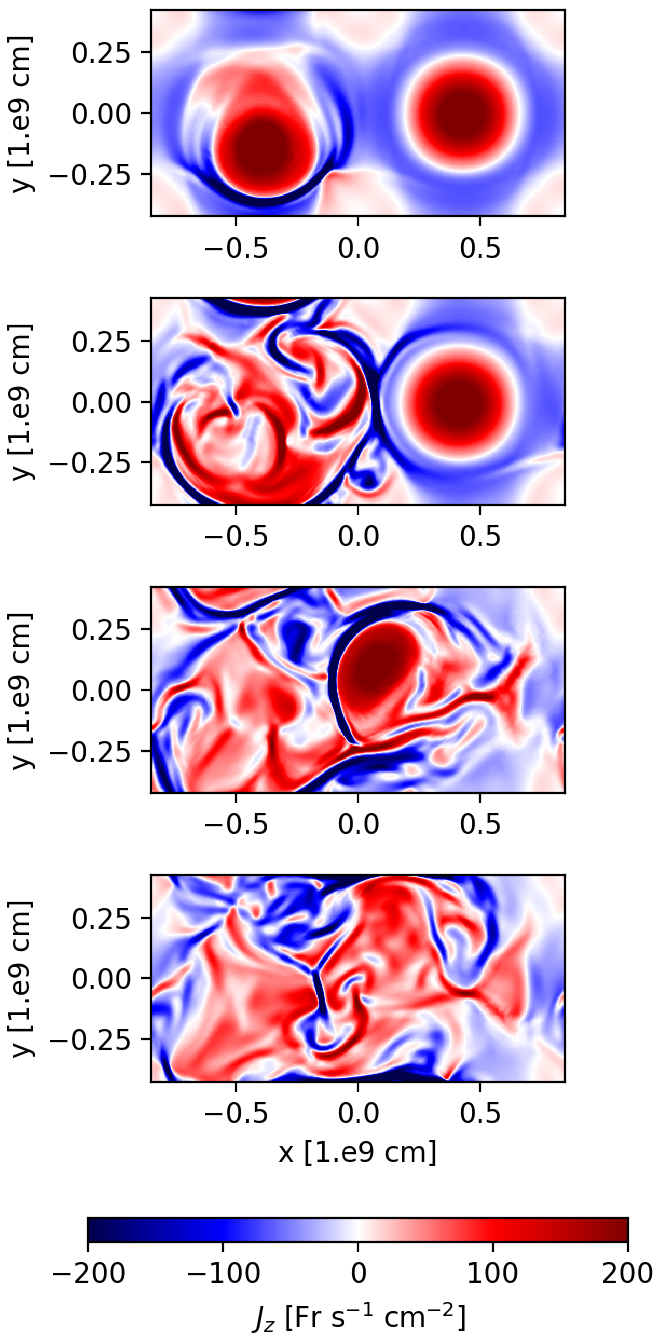}
  \includegraphics[width=0.3\hsize]{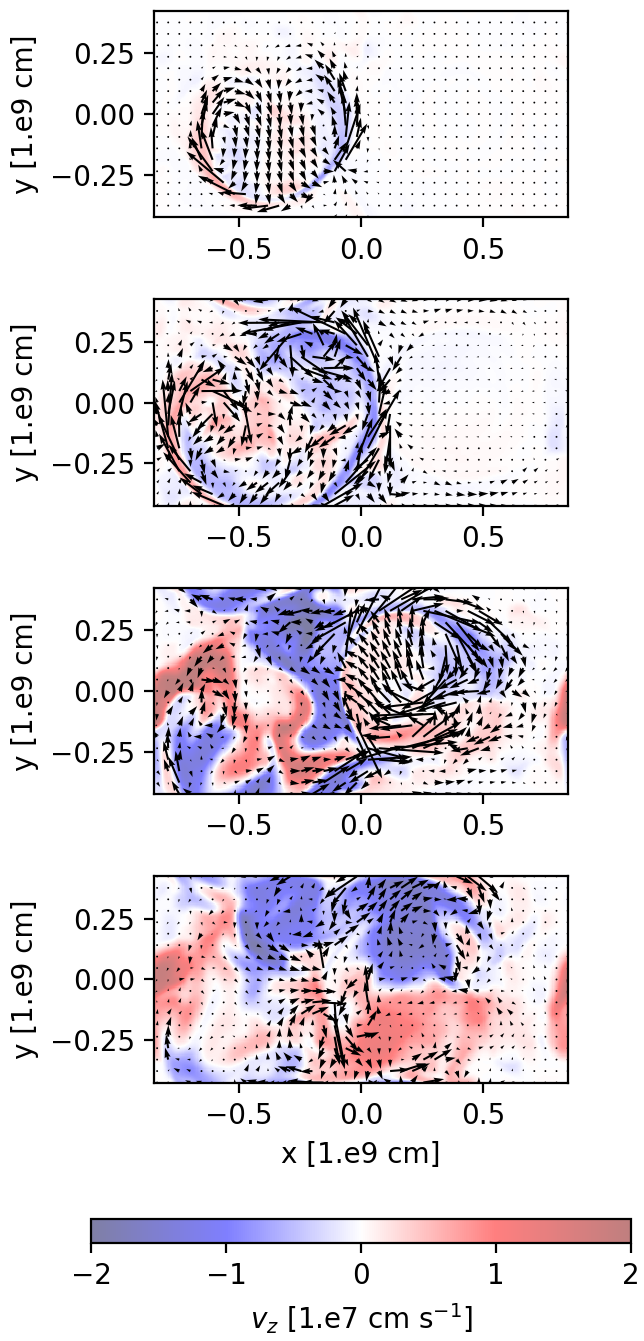}
    \includegraphics[width=0.3\hsize]{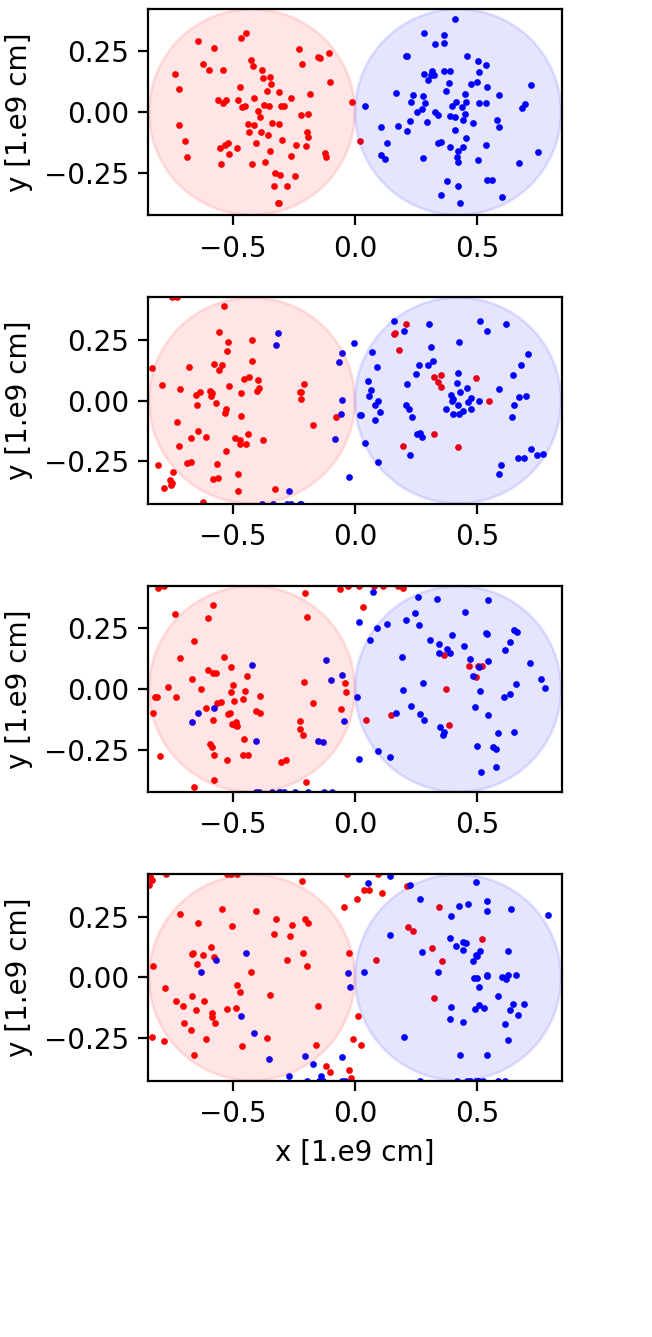}
  \caption{\textbf{First column}. Horizontal cut of the current density across the mid plane at times (from the top) $t = 12400\,\mathrm{s}$ (onset of first kink instability), $12500\,\mathrm{s}$, $12550\,\mathrm{s}$ (second strand’s disruption), and $12600\,\mathrm{s}$. \textbf{Second column}. Horizontal cut of the velocity across the mid plane at the same 4 times. Arrows show the orientation of the velocity field. Color-maps evaluate the intensity of the vertical component of the velocity field. \textbf{Third column}. Terminal locations ($z = L$) of sample field lines at the same 4 times. Red field lines (spots) depart from the $z= -L$ footpoint on the left, (shaded red region), and blue field lines from the the right (blue shaded region).
  Initially, red and blue field lines are randomly distributed inside the blue and the red circles, respectively. 
  Subsequent starting locations at the lower boundaries points are determined at later times by tracking their locations in response to the} photospheric motions.
  \label{Fig:RESULTS_current}
\end{figure*}

\begin{figure}[h!]
\centering
  \includegraphics[width=0.92\hsize]{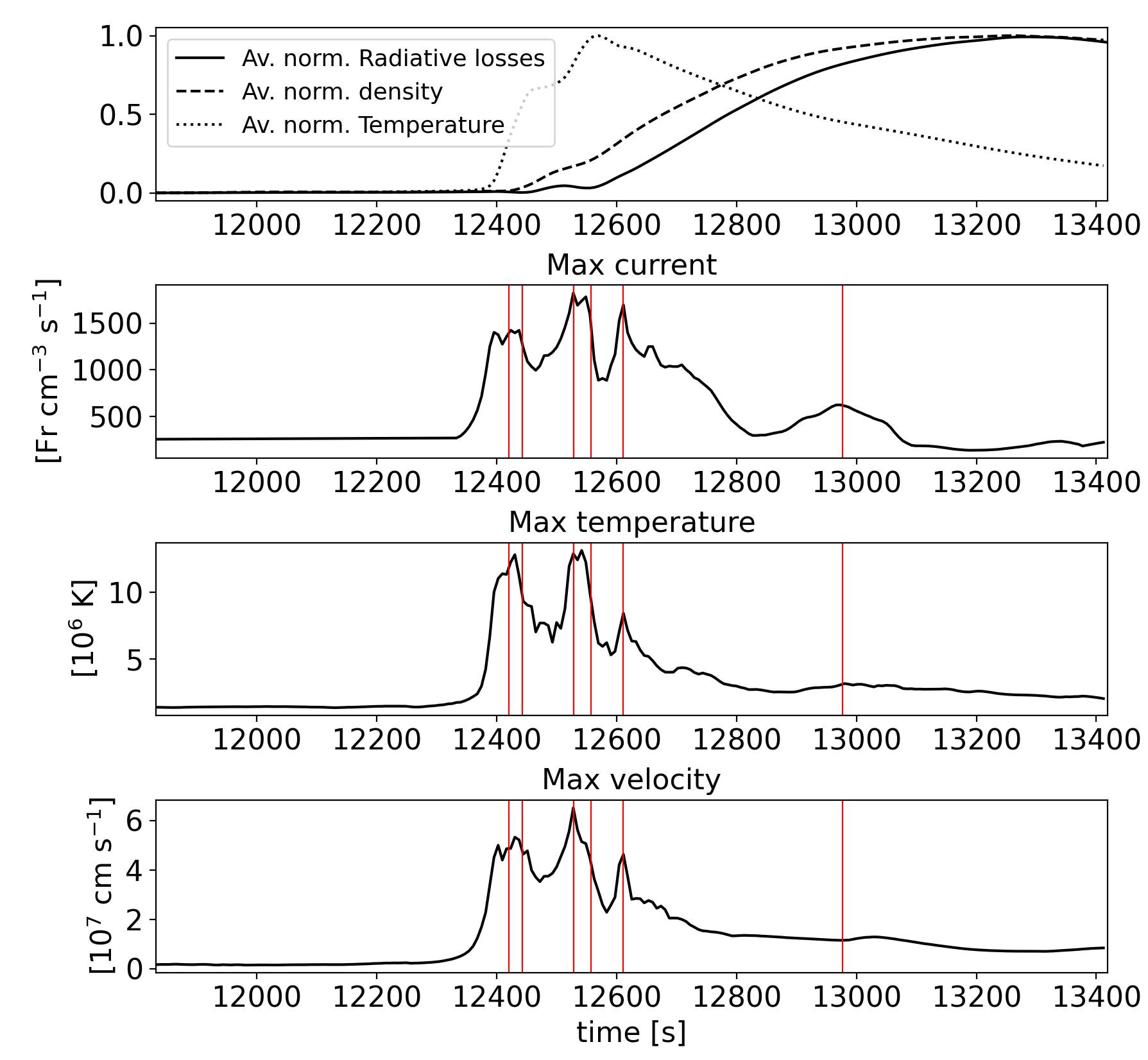}
  \caption{From top to bottom: average, normalized radiative losses, density and temperature in the corona, maximum current density, maximum temperature, and maximum velocity vs. time. Red vertical lines mark the times of large heating events.}
  \label{Fig:RESULTS_max_values}
\end{figure}
\begin{figure*}[h!]
\centering
  \includegraphics[width=\hsize]{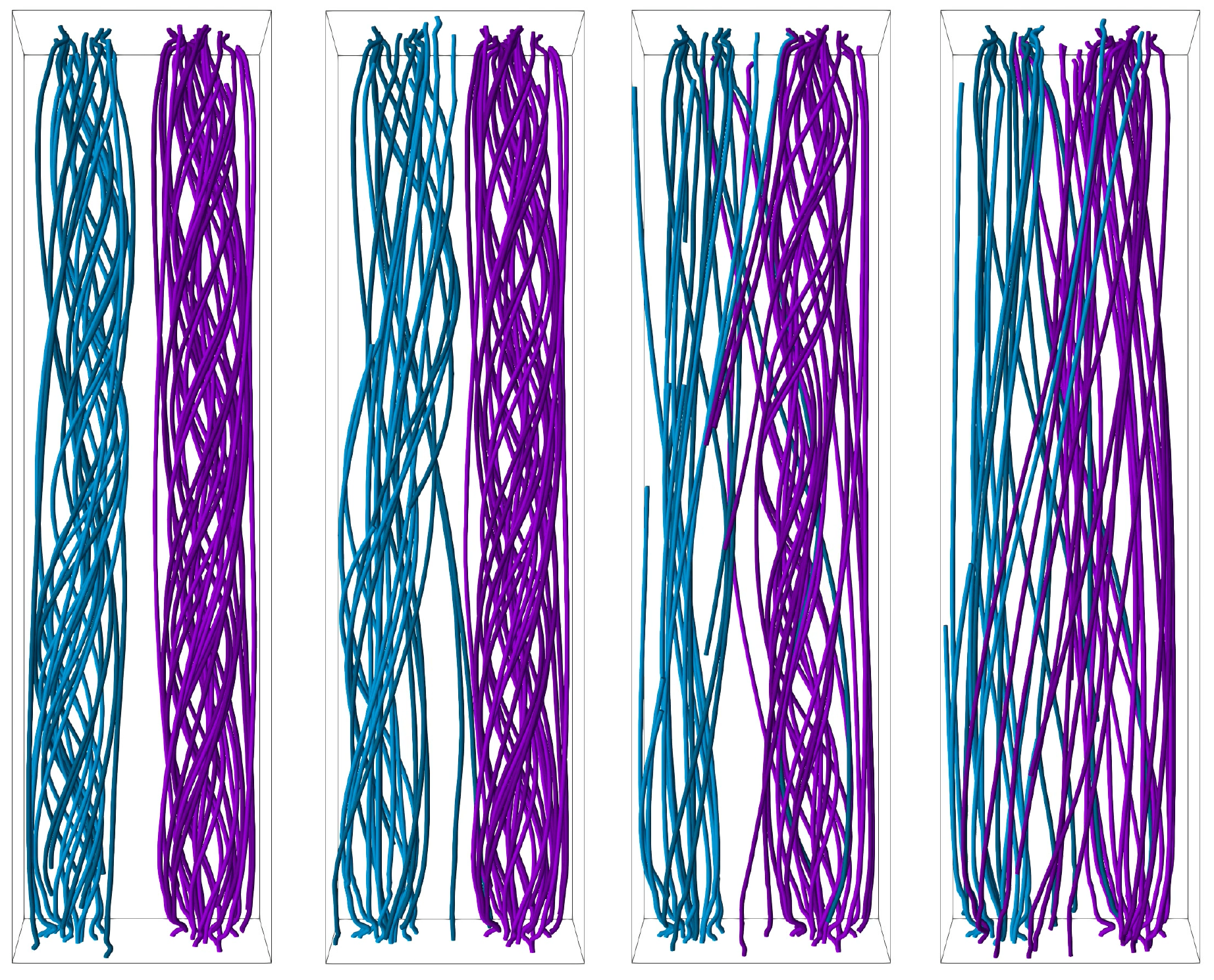}
  \caption{3D rendering of the magnetic field lines in the box around the two flux tubes at times:  (from the left) $t = 12400\,\mathrm{s}$ (onset of first kink instability), $12500\,\mathrm{s}$, $12550\,\mathrm{s}$ (second loop disruption), and $12600\,\mathrm{s}$. The change in the field line connectivity during the evolution of the MHD cascade is highlighted by the different colors.}
  \label{Fig:RESULTS_3D_evolution}
\end{figure*}
\begin{figure}[h!]
\centering
  \includegraphics[width=\hsize]{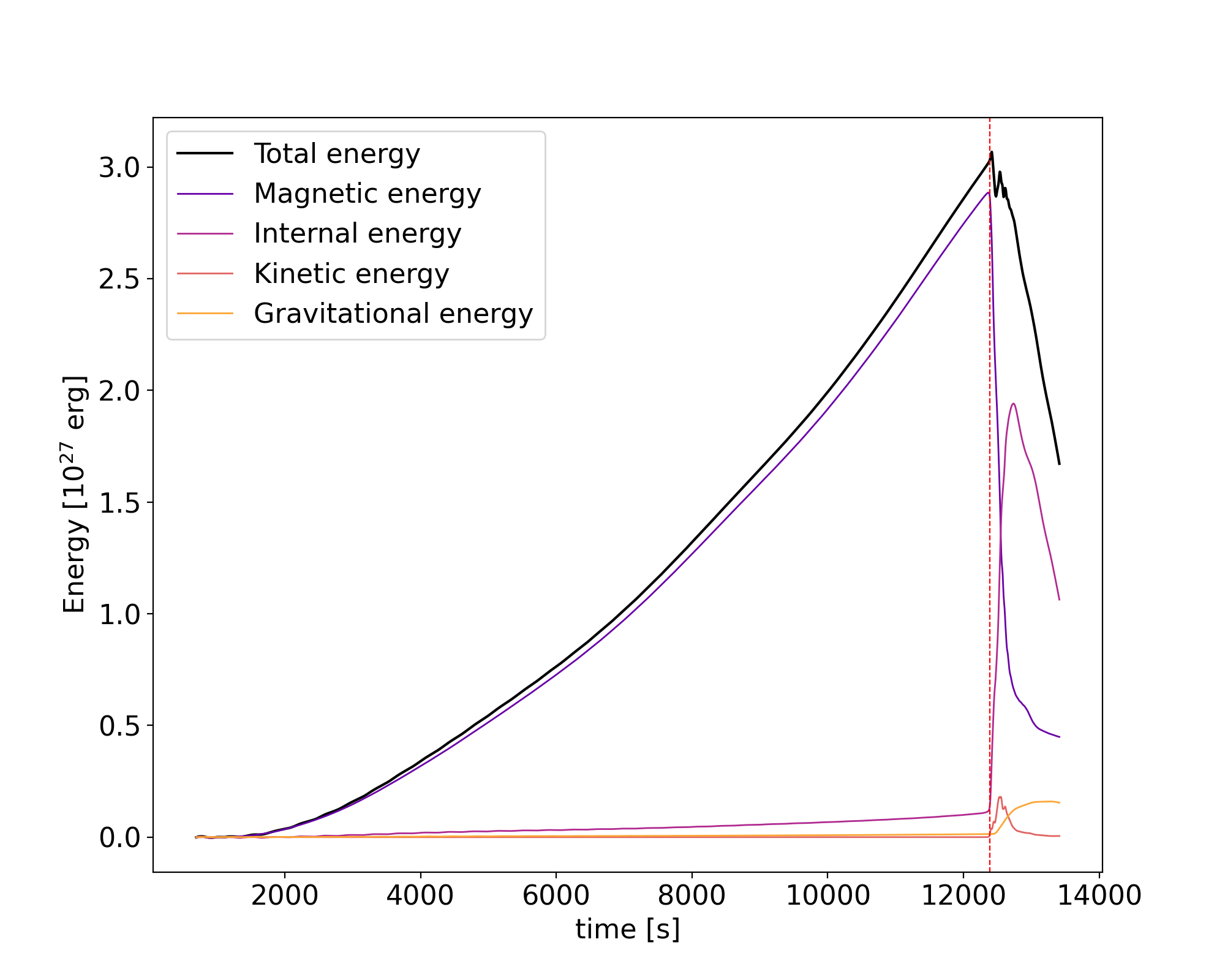}
  \caption{Magnetic (\textbf{purple curve}), internal (\textbf{pink curve}), kinetic (\textbf{orange curve}), and gravitational (\textbf{yellow curve}) energies, as functions of time. The \textbf{black curve} is the total energy given as the sum of the four energy terms. The onset of the instability is marked (\textbf{dotted red vertical line}).}
  \label{Fig:RESULTS_total_energy}
\end{figure}
\begin{figure}[h!]
\centering
  \includegraphics[width=\hsize]{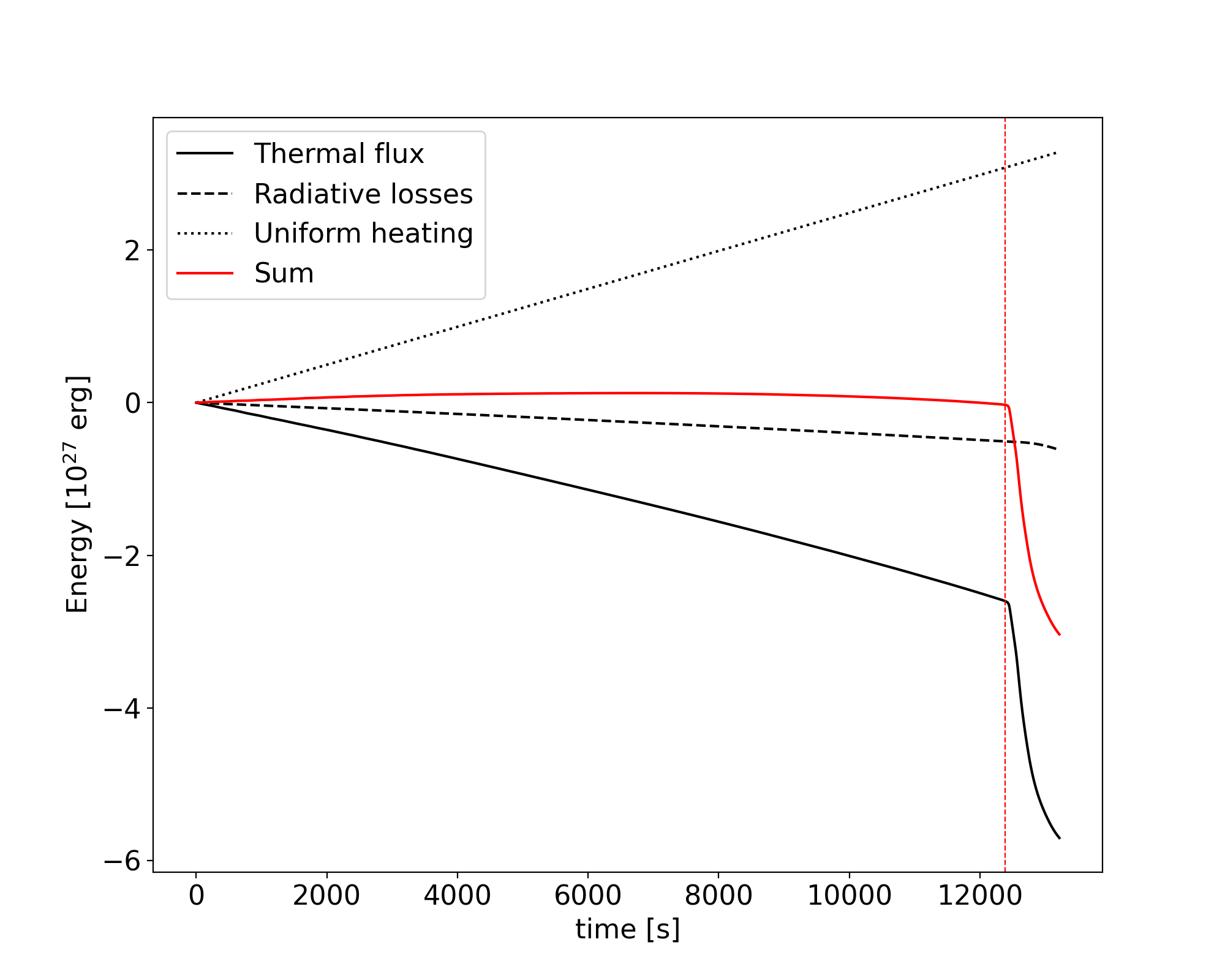}
  \caption{Time-integrated thermal flux (\textbf{solid black curve}), radiative losses (\textbf{dashed black curve}), and background heating (\textbf{dotted, black curve}) as functions of time. The \textbf{solid, red curve} is the sum of the three contributions. The onset of the instability is marked (\textbf{dotted red vertical line}).}
  \label{Fig:RESULTS_losses}
\end{figure}
\begin{figure}[h!]      
\centering
  \includegraphics[width=\hsize]{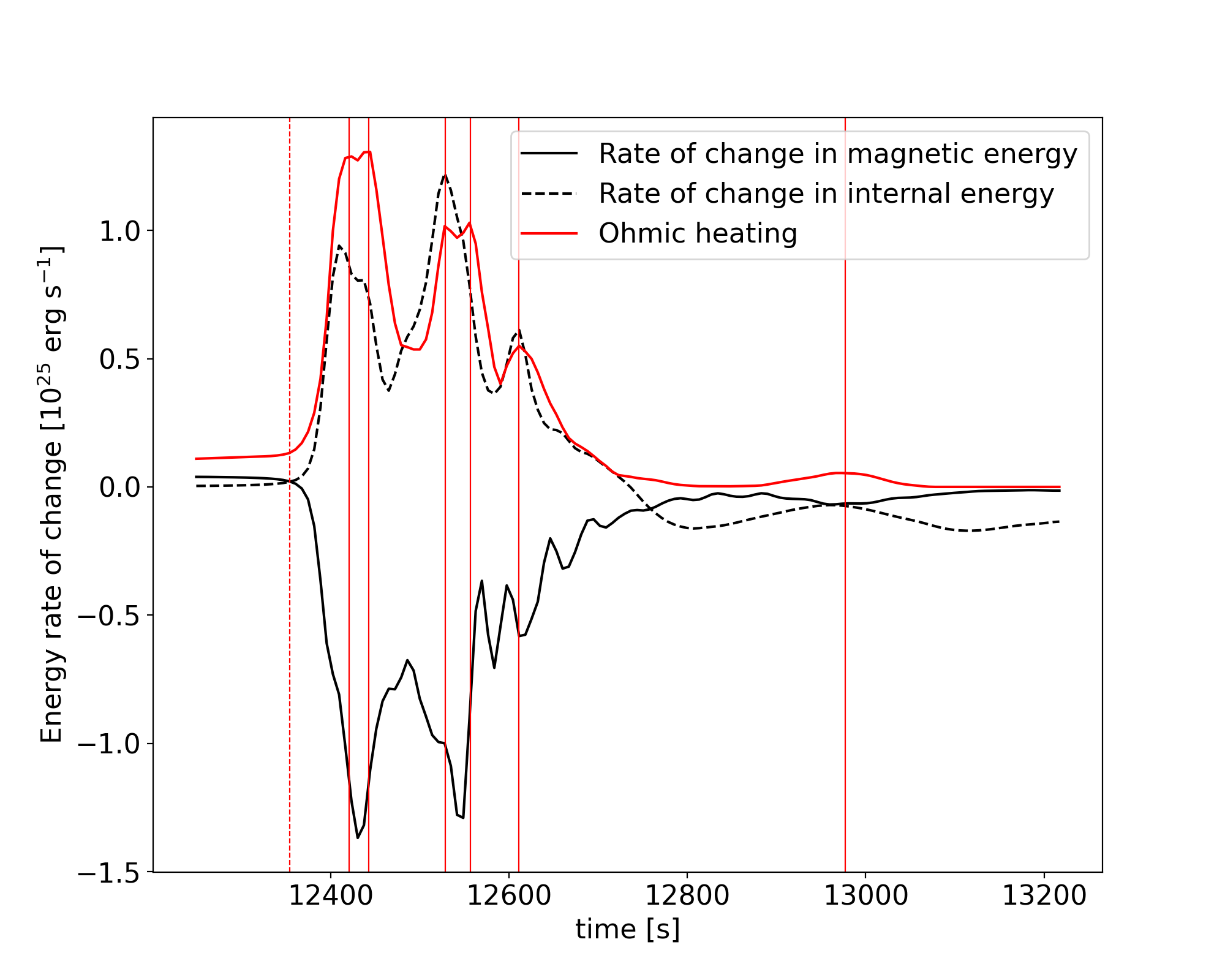}
  \caption{Rates of change in magnetic (\textbf{solid black curve}) and internal (\textbf{dashed black curve}) energies, and Ohmic heating (\textbf{solid red curve}), as functions of time. Solid, red vertical lines highlight times of large heating events. The onset of the instability is marked (\textbf{dashed red vertical line}).}
  \label{Fig:RESULTS_magnetic_rate}
 \end{figure}
\begin{figure}[h!]
\centering
  \includegraphics[width=\hsize]{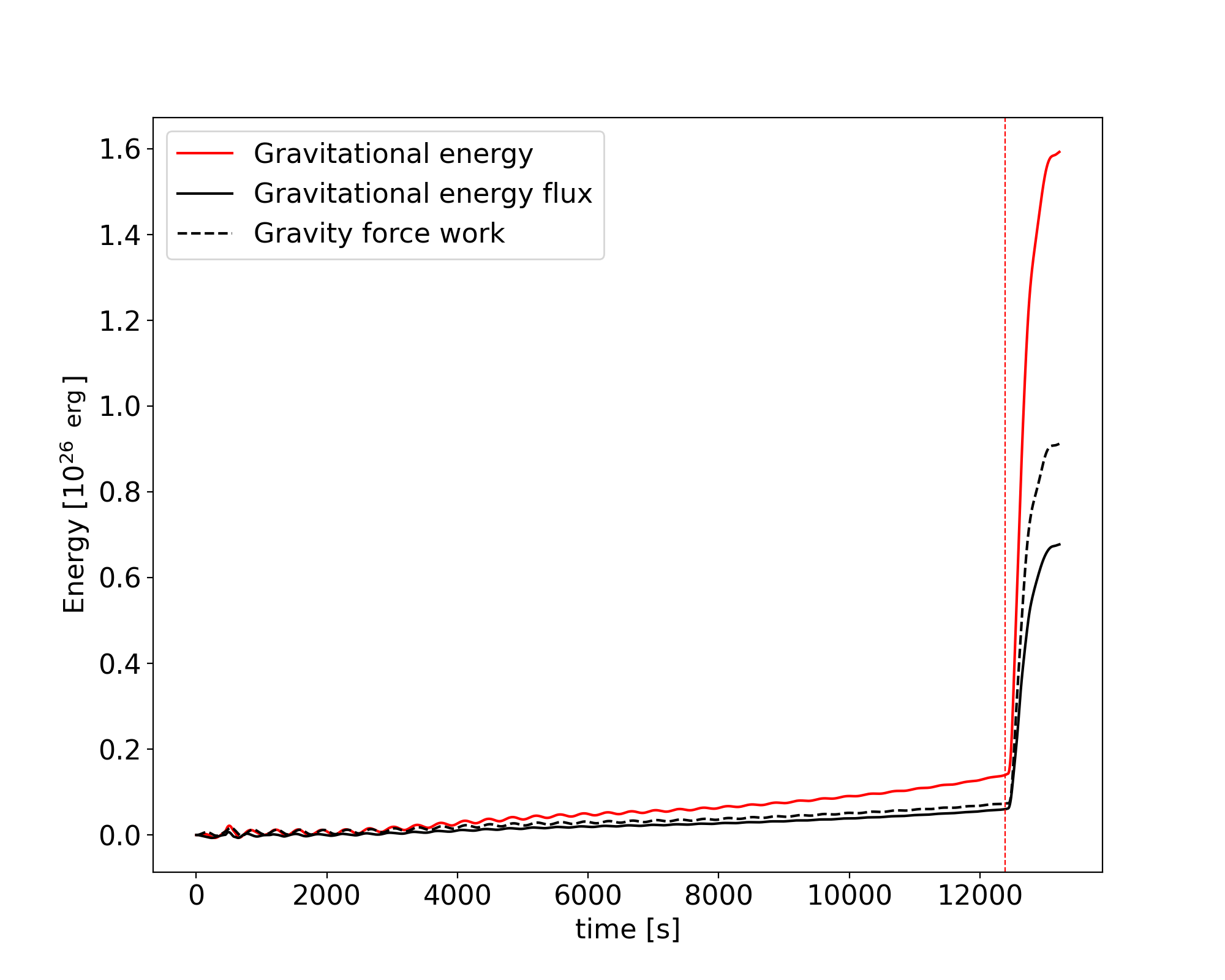}
  \caption{Gravitational energy \textbf{solid, red curve}, time-integrated gravitational energy flux (solid black curve), and work done by gravity (dashed black curve) as functions of time.   The onset of the instability is marked (\textbf{dashed red vertical line}).}
  \label{Fig:RESULTS_gravitational_energy}
\end{figure}
\begin{figure*}[h!]
\centering
  \includegraphics[width=\hsize]{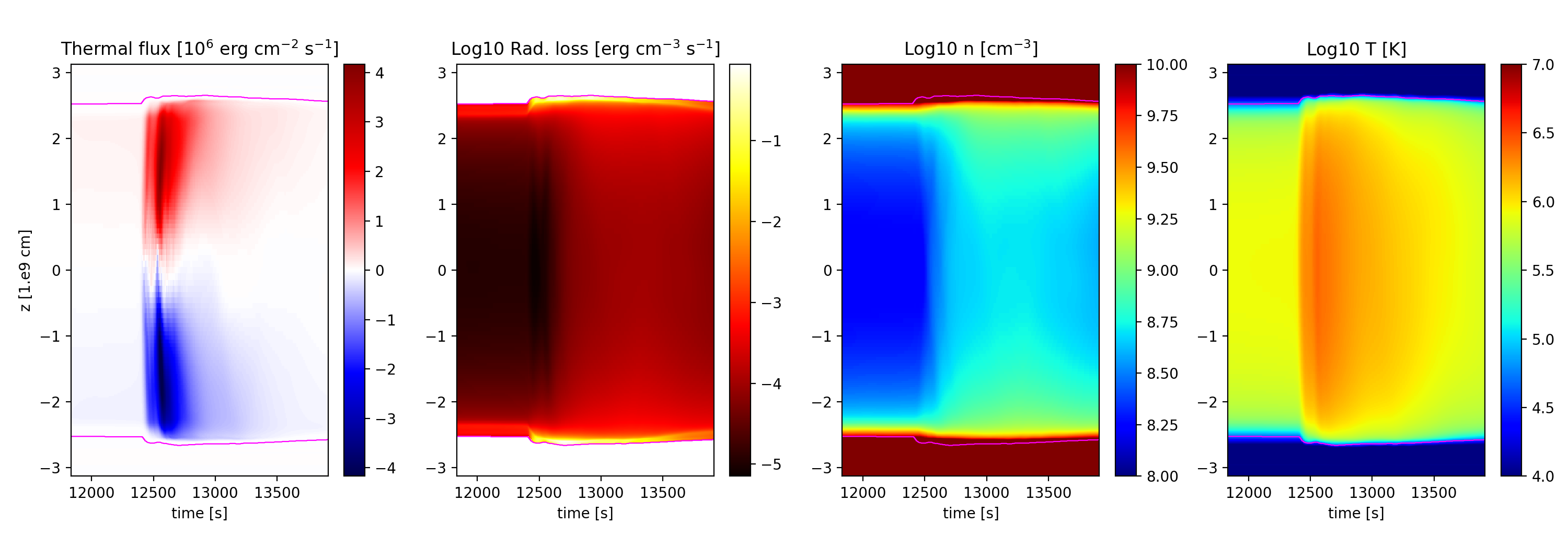}
  \caption{Average vertical thermal flux, radiative losses, plasma density and temperature vs. time (on the horizontal axis) and height (z; on the vertical axis). Averaging is on the horizontal planes. The region of the domain where the temperature exceeds $10^4\,K$ (i.e. transition region and corona) is bounded (magenta lines).}
  \label{Fig:RESULTS_zt_plot}
\end{figure*}
\begin{figure*}[h!]
\centering
\hspace{0.8 cm}
\includegraphics[height=0.65\hsize]{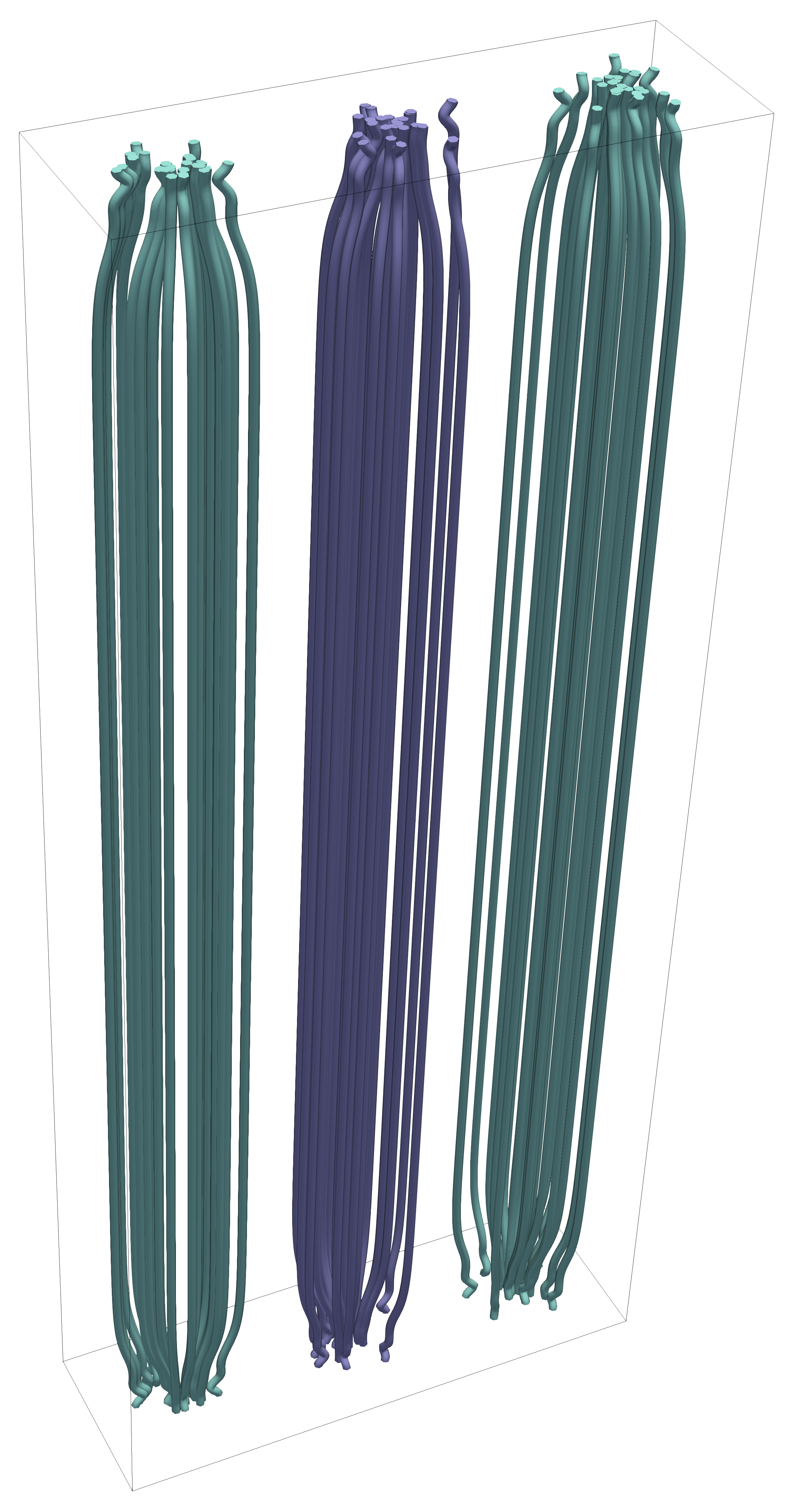} \hspace{1.4 cm}
  \includegraphics[height=0.65\hsize]{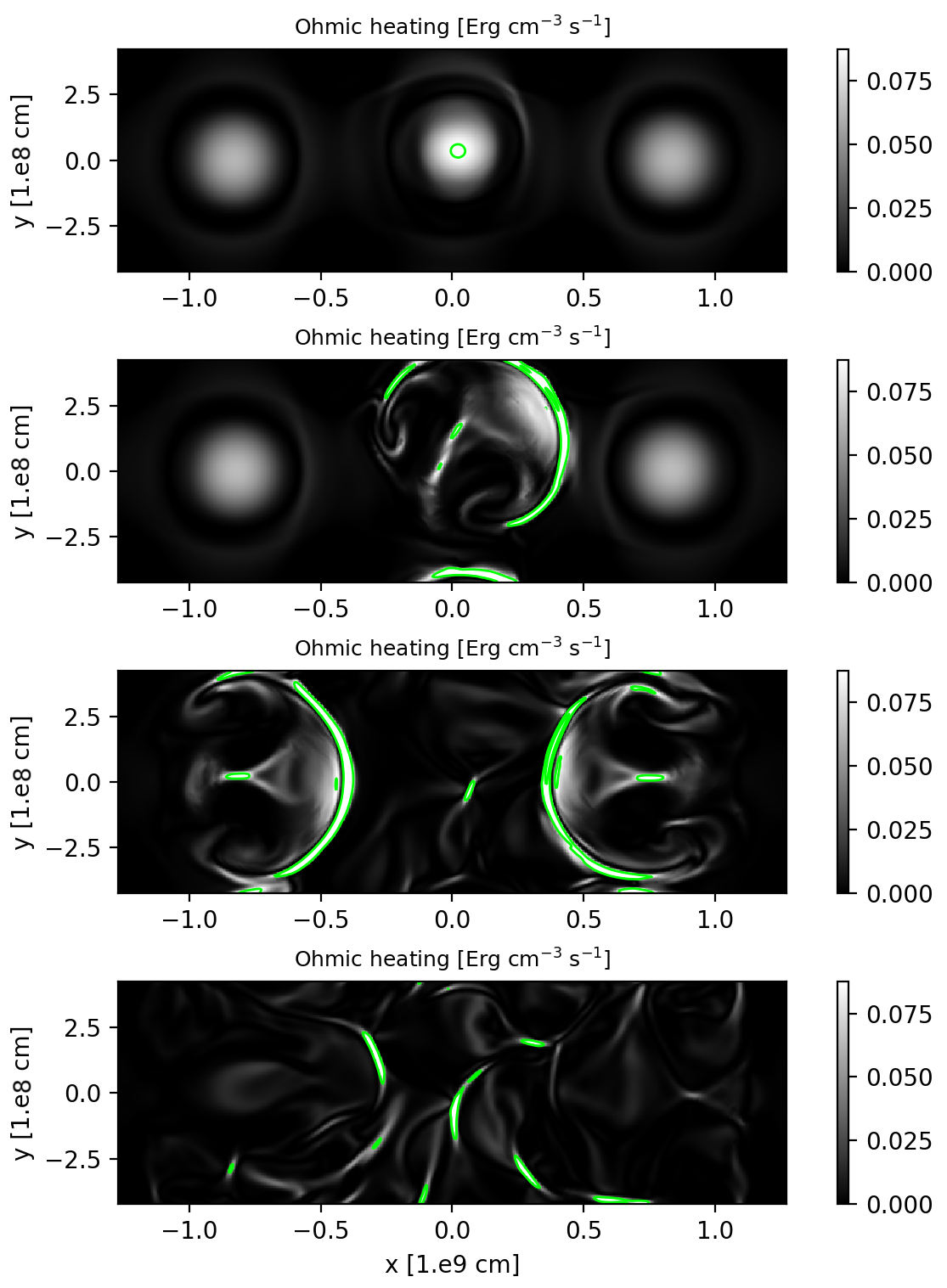}
  \caption{\textbf{Left panel}: 3D rendering of the initial magnetic field configuration in the proximity of each coronal loop (for the second, three-strands loop model). Purple field curves will be subjected to a faster twisting driver than the green ones. \textbf{Right panel}: horizontal cut of the Ohmic heating per unit time and unit volume at the middle of the box at times (from the top) $t = 11200\,\mathrm{s}$ (onset of first kink instability), $11400\,\mathrm{s}$, $11800\,\mathrm{s}$ (second and third loops' disruption), $11900\,\mathrm{s}$. 
  In the green filaments the current density exceeds the threshold value for dissipation.}
  \label{Fig:RESULTS_three_loops}
\end{figure*}
\subsection{Continued driving: evolution before the instability}
The box size is 6.2 Mm in the $z$ direction. The chromosphere extends for 0.7 Mm on both sides, and the corona (including the transition region) is in the middle $2L = 5 \times 10^9\,\mathrm{cm}$ (see sec. \ref{sec:thermal}).  
Here we safely restrict our analysis to the inner domain 
between $z = \pm 2 \times 10^9\,\mathrm{cm}$ in order to avoid possible undesired contributions as a result of expected changes in transition region height. Since boundary conditions are periodic at the side boundaries of the box, fluxes are only evaluated at the upper and lower boundaries of the sub-domain i.e. at $z = \pm 2 \times 10^9\,\mathrm{cm}$.

Initially, the two flux tubes are slowly twisted, at a speed much slower than the Alfvén speed. As a consequence, the initial evolution of the magnetic structure is in a quasi steady-state. In particular, an azimuthal magnetic field component grows almost linearly with time.
The magnetic torsion is transmitted to the coronal part of the magnetic tube (i.e. at $|z| < 2 \times 10^9\,\mathrm{cm}$) after two hundred seconds, in accordance with the time estimated for a magnetic signal to cross the chromospheric layer. 

Figure \ref{Fig:RESULTS_rate_of_change} shows the rate of change of the total energy, which is given as the sum of magnetic, kinetic, thermal and gravitational energy.
Total energy is not constant inside the coronal volume as a result of incoming fluxes at the chromospheric boundaries of the domain (such as Poynting flux, kinetic energy flux, enthalpy flux, gravitational energy flux, and thermal conduction), energy sources (background heating), and sinks (radiative losses). Total energy plus incoming/outgoing energy (see Eq. \ref{eq:energy}) is approximately conserved throughout the numerical experiment. 
Among all the external contributions, the Poynting flux is dominant during the build-up of the twisting.

The initial evolution of the system might be seen as the superposition of a long-lasting and steady tube twisting (where magnetic energy and current density slowly grow as a consequence of the field lines torsion) and a wave-like response to the induced dynamics (where oscillations of short characteristic timescales are damped with time).
In particular, long period oscillation ($P \approx 360\,\mathrm{s}$) are clearly visible in Fig.~\ref{Fig:RESULTS_rate_of_change} and in Fig.~\ref{Fig:RESULTS_BV_frequency} which shows the pressure force work and the gravity force work rates as function of time. Those features might be associated with Brunt–Väisälä oscillations whose characteristic frequency is:
\begin{equation}
    \omega = \sqrt{- \frac{g}{\rho} \parder{\rho (h)}{h}},
\end{equation}
where $h (z) = \frac{2 L}{\pi} \cos{\left(\frac{\pi z}{2 L} \right)}$ gives the height in a semi-circular loop.
In particular, a frequency of $2.76 \times 10^{-3}\,\mathrm{s}^{-1}$ (corresponding to period of approximately $360\,\mathrm{s}$) matches the theoretical value at transition region heights, suggesting that the nature of this oscillation might be associated with buoyancy movements in the upper chormospheric layers.
Alfvén waves appear in each thread as azimuthal modes with period of nearly $50\,\mathrm{s}$.

In Fig.~\ref{Fig:RESULTS_kinetic_energy}, the kinetic energy reaches a steady state value around a time of 6000 seconds and remains there until about 11000 seconds when it exponentially increases as the first kink instability occurs. 
The theoretical limit, computed from Eq. (\ref{eq:kin_energy}) and shown as a red dashed line, agrees with the actual volume-averaged kinetic energy. 
Oscillations in kinetic energy have a period of approximately $180\,\mathrm{s}$ (see. Fig.~\ref{Fig:RESULTS_BV_frequency}) and are the result of magnetosonic waves.
Moreover, the growth of the kinetic energy during the onset of the \textit{ideal} instability is, as expected, exponential with time. This is shown in the internal panel of Fig.~\ref{Fig:RESULTS_kinetic_energy}. In particular, the slope of the exponential increase matches with the theoretical value $\tau = 0.1 \times 2L/v_A$, where $v_A$ is the Alfvén velocity \citep{van1999complete, hood2009coronal}.

Fig.~\ref{Fig:RESULTS_magnetic_energy} shows that the magnetic energy determined from the simulation is very close to the prediction of the theoretical model presented in the previous section. The quadratic increase of the magnetic energy is determined by the linear growth of the azimuthal component of the magnetic field during the twisting. 

The vertical component of the current density dominates over the other ones. It also grows linearly in time as a consequence of the magnetic tube twisting (see Fig.~\ref{Fig:RESULTS_max_current}).
As assumed in Eq. (\ref{eq:max_current}), the maximum current intensity is along the axis of each flux tube  As shown in Fig.~\ref{Fig:RESULTS_max_current_2}, the axial current remains positive around the centres of the strands. On the outer edge, there is a neutralizing negative current ensuring the net axial current remains zero.

\subsection{Onset of the instability}

The first tube becomes unstable after around $12400\,\mathrm{s}$. We estimated the amount of twist, defined as $\Phi = 2 \pi N$ (with $N$ the number of twist turns in the unstable strand), at the time of the instability. We considered both the maximum tangential photospheric velocity $v_{\phi}$ and an averaged value $\langle v \rangle = \int_0^{2 r_{\mathrm{max.}}} v_{\phi} r dr / \int_0^{2 r_{\mathrm{max.}}} r dr$. In the first case, the $\Phi \sim 10$ while in the second one it is a factor two smaller. 
In both cases, $\Phi$ is of the same order of magnitude as previous results such as the Kruskal-Shafranov condition  ($\Phi_{\mathrm{KS}} = 3.3\,\pi$) \citep{priest2014magnetohydrodynamics}.

The onset of the first kink instability and the subsequent MHD cascade can be followed by inspecting the current density and velocity evolution. For instance, Fig.~\ref{Fig:RESULTS_current} shows the current density distribution (first column) and the velocity field (second column) over the loop mid plane at four different times. In the first panel, the onset of the first kink instability is shown: the unstable flux tube begins to flex as a consequence of the growing magnetic pressure imbalance. As a consequence of that, a single current sheet forms at the edge of the loop and the velocity grows at its sides. Subsequently, in the second panel, the current sheet fragments in a turbulent way (see the velocity map) into smaller current sheets and the entire structure expands to interact with the neighbouring loop. That causes the second loop's instability. The third and fourth panels show the evolution of the MHD cascade (i.e. second loop disruption) triggered by the first instability.
All the time, zones of high plasma velocity on the horizontal mid-plane spread over regions of high current density. This is expected since plasma is mostly accelerated by magnetic forces where magnetic field gradients are higher.

The average temperature peaks $100\,\mathrm{s}$ after the onset of the instability, while average density and radiative losses reach the maximum value after other $800\,\mathrm{s}$ (see first panel of Fig.~\ref{Fig:RESULTS_max_values}).
The whole, turbulent evolution of the system is harder to follow, but quantitative information on its dynamics can be obtained from the maximum current, temperature and velocity evolution shown in Fig.~\ref{Fig:RESULTS_max_values}. The three plots show the same qualitative behaviour with some high peaks around $t = 12500\,\mathrm{s}$, during large heating events corresponding to dissipation of relatively large current sheets. In particular, the first group of peaks occurs during the onset of the first kink instability. The second and third ones correspond to the times when the second loop is destabilized and when it is finally disrupted, respectively. Another peak in the current intensity occurs at $t = 13000\,\mathrm{s}$, followed by a moderate enhancement in the loop temperature and velocity. It is produced by the formation and subsequent dissipation of a big current sheet induced by the continuous driving at the boundaries.

The dissipation of the multi-threaded loop into smaller current sheets can be traced by following the magnetic field lines connectivity over the time. The third column of Fig.~\ref{Fig:RESULTS_current} shows the end points of some field lines on the upper boundary plane $z = z_{\mathrm{max.}}$. Red points are connected at the opposite side (i.e. $z=-z_{\mathrm{max.}}$) to the footpoint of the left. \emph{Vice versa}, blue dots refer to field lines connected to the right-hand footpoint. 
Field lines are traced from the bottom side of the box and mapped into the upper one using a second order Runge-Kutta integration scheme.
The location of the starting points at $z=-z_{\mathrm{max.}}$ are updated according to the imposed rotation, while the points at the opposite side are expected to change as the field lines move or change by reconnection.
It is easy to see that the field line connectivity changes as soon as the MHD cascade takes place and that magnetic reconnection has occurred in the meantime. Indeed, during the instability, field lines from each strand become entangled and eventually cross the lateral boundaries of the domain. The same thing is likewise evident in Fig. \ref{Fig:RESULTS_3D_evolution} where field lines in the box are shown in full-3D rendering. Field lines are computed using a fourth-order Runge-Kutta scheme and  color is attributed depending on whether the starting points in the photosphere are placed. As the twisting triggers the kink instability, field lines reconnect with each other. At the end of the process, some light blue and purple lines connect different loop footpoints.

The energetics of the numerical experiment reflect the physical processes that drive the system dynamics.
Fig.~\ref{Fig:RESULTS_total_energy} shows the evolution of the four energy components i.e. magnetic energy, kinetic energy, thermal energy and gravitational energy. The magnetic energy dominates over the other components for all the initial, smooth evolution of the system. As mentioned above, the main source of energy derives from the Poynting flux (see Fig.~\ref{Fig:RESULTS_rate_of_change}). The net effect of thermal conduction, radiative losses and background heating is negligible provided that the magnetic field changes are slow compared with the radiative and conductive time scales. In Fig.~\ref{Fig:RESULTS_losses}, the sum of the time-integrated thermal flux, radiative losses and uniform heating is practically zero while the thermal conduction dominates over the radiative losses, as expected in typical coronal conditions.

After the onset of the MHD avalanche, the magnetic energy rapidly drops. The kinetic energy increases exponentially but remains at least one order of magnitude smaller than the magnetic energy.  Most of the magnetic energy gained, through footpoint driving, is converted into heat. The steep rise in thermal energy follows the plasma acceleration. In particular, Fig.~\ref{Fig:RESULTS_magnetic_rate} shows how the rate of change in magnetic energy matches the instantaneous Ohmic heating. In turn, it influences the rate of change in heating.

The several heat pulses released after the multiple magnetic reconnection events enhance the thermal conductive flux towards both transition regions (see Fig.~\ref{Fig:RESULTS_losses}). 
The heat flow is then slowed down in the chromosphere because conduction is less efficient at cooler temperatures. As a consequence of that, an excess of pressure builds up in the transition region and the top of the chromosphere. This creates the pressure gradient that causes the evaporative up-flow. The plasma expanding upwards, in turn, leads to an increase in the coronal density inside the magnetic structure. The sudden growth of the gravitational potential energy traces this strong mass flow upwards, as shown in Fig.~\ref{Fig:RESULTS_gravitational_energy}.
After the beginning of the MHD avalanche, the gravitational energy increases as a consequence of the chromospheric plasma evaporation in the coronal volume. It supplies gravitational energy flux at the boundaries while the remaining contribution to the potential energy is given by the work done by the gravity force to distribute this denser plasma over the entire loop length.

Fig.~\ref{Fig:RESULTS_zt_plot} shows the average vertical thermal flux, radiative losses, density and temperature as a function of time and height. The strongest thermal flux (first panel) develops at the times when each loop is disrupted, i.e., when the temperature gradient it the greatest. The heat flux propagates toward the upper and lower transition regions and is stronger in the corona. On the other hand, radiative losses (second panel) are stronger at later times when the density (third panel) has increased by chromospheric evaporation. The biggest contribution is localized in the transition region where the rates exceed the coronal radiative losses by at least two orders of magnitude. As heating released during the instability is rapidly spread (in few tens of seconds) along the tube, temperature (fourth panel) uniformly rises. It then slowly decreases from $10\,\mathrm{MK}$ to $1\,\mathrm{MK}$ on a timescale of $1000\,\mathrm{s}$.

\subsection{Three-loop simulation}
\label{sec:avalanche}

The propagation of the instability described so far is an avalanche process which can extend to more and more nearby flux tubes \citep{2016mhhoodd}. 
To ensure this, we performed a second numerical experiment with three interacting coronal loops strands. The initial configuration of the magnetic structure is shown on the left panel of figure Fig.~\ref{Fig:RESULTS_three_loops}. As in the previous case, this magnetic structure is embedded in stratified atmosphere including a cold ($T \simeq 10^4 \mathrm{K}$ ) chromospheric layer and a hot and tenuous corona ($T \simeq 10^6 \mathrm{K}$ and $n \simeq 10^9\mathrm{cm}^{-3}$). Eqs. (\ref{eq:mass_conservation})-(\ref{Eq:electric_field}) summarize the underlying physics driving the evolution of the system, as discussed in Sec. \ref{sec:model}. Similarly to the first case, one magnetic strand is twisted faster than the other ones at its footpoints and become kink-unstable. 
For instance, the right panel of Fig.~\ref{Fig:RESULTS_three_loops} shows the propagation of the instability of a central faster tube to two adjacent tubes. As expected, the most unstable loop triggers the global dissipation of the magnetic structure into smaller current sheets.
Heating by Ohmic dissipation is localized inside relatively small regions where the current density is higher. These current sheets develop and spread in a turbulent way throughout the entire extension of the domain.

\section{Discussion and conclusions}
\label{sec:conclusions}

This work addresses the energy released impulsively in the corona under strong magnetic stresses. In particular, we have shown that MHD avalanches are efficient mechanisms for fast release of magnetic energy in the solar corona progressively stored by slow, uniform photospheric motions.
We describe a system consisting of two neighbouring twisted flux tubes. These interacting flux tubes comprise a stratified atmosphere, including chromospheric layers, a thin transition region to the corona, and the associated transition from high-$\beta$ to low-$\beta$ regions.
The model includes the effects of thermal conduction and of optically thin radiation.
Rotation of the plasma at the upper and lower sides of the box applies twisting to the magnetic flux tubes. Since line-tying of the field lines at the photospheric boundaries is expected to be maintained over time by high plasma beta values and a sufficient spatial resolution, each loop can develop high levels of twist, which increase the azimuthal component of the magnetic field. Above a certain stress threshold, the structure becomes kink-unstable and suddenly relaxes to a new equilibrium configuration \citep{hood2009coronal}. In particular, since one strand is twisted faster than the other, it will become unstable before the other and trigger the avalanche process that, in turn, will spread as it affects the neighbouring flux tube.
Magnetic reconnection between unstable flux tubes causes bursty and diffuse energy release (similar to a nanoflare storm) and changes the field connectivity. Moreover, through repeated reconnection events, the system relaxes towards the minimum energy state. 
The system undergoes an initial dynamic phase where the plasma is rapidly accelerated. The initial helical current sheet progressively fragments in a turbulent way into smaller current sheets which, in turn, dissipate magnetic energy via Ohmic heating. As soon as the steep rise in kinetic energy is damped in the corona, the released heating rapidly increases coronal temperatures and, consequently, the pressure scale height. As a consequence of that, the steep rise in temperature is followed by a progressive coronal density enhancement as a result of chromospheric evaporation.

Results achieved in this paper agree with those found by \cite{hood2009coronal}, \cite{2016mhhoodd} and \cite{reid2018coronal}, but go further and extend them.
In particular, we demonstrate that even inside a stratified atmosphere, highly twisted loops with zero net current undergo the non-linear phase of the kink instability where reconnection in a single current sheet triggers the fragmentation of the flux tubes at multiple reconnection sites.
In particular, once the first unstable strand is disrupted, it coalesces with the neighbouring strands inducing an MHD cascade, as found in uniform coronal atmospheres \citep{tam2015coronal, 2016mhhoodd, reid2018coronal}.

 As shown in Fig.~\ref{Fig:RESULTS_magnetic_rate}, magnetic energy is released in discrete bursts as stable strands are disrupted and single current sheets are dissipated. This bursty heating does not show evidence of reaching a steady state. 

Thermal conduction is very effective in spreading heating along field lines, and this leads to the filamentary structuring in loop temperature. Also, as shown in Fig.~\ref{Fig:RESULTS_max_values}, because of thermal conduction, the temperature grows to about $10^7\,\mathrm{K}$, much less than around $10^8\,\mathrm{K}$ found in \cite{hood2009coronal}. Peak temperatures of few $10^7\,\mathrm{K}$ as well as variations in magnetic and internal energy of $10^{27}\,\mathrm{erg}$ agree with those found from microflares observations \citep{testa2020coronal}.

Radiation also has an important effect in the temperature distribution. Radiative losses are stronger across the transition region, where the plasma density is higher. As the upper atmosphere is heated, this layer acts as a thermostat for the corona since it tends to restore the initial coronal temperature. Indeed, it maintains the temperature gradient that allows heat to flow out of the corona.

A deep chromospheric layer is important to guarantee line-tying throughout the whole evolution of the coronal loop. Then, photospheric motions can slowly twist the magnetic flux tubes expanding across the chromospheric layer. At the same time, the chromosphere acts as reservoir of dense plasma which can flow into the corona, as a consequence of impulsive heating.  Modelling the chromospheric evaporation and the resulting increase of the loop emissivity is fundamental to corroborate the results through comparison with EUV and X-ray observations of dynamic coronal loops.

As shown is section \ref{sec:avalanche}, the propagation of the instability is an avalanche process which can extend to more and more nearby flux tubes \citep{2016mhhoodd}. 

In conclusion, this work confirms, and constrains the conditions for, the propagation of a kink instability among a cluster of flux tubes, including a more complete, stratified loop atmosphere, and important physical effects, in particular the thermal conduction and the optically thin radiative losses. The avalanche can trigger the ignition and heating of a large scale coronal loop, with parameters not far from those inferred from the observations. 

The detection of the small scales involved will be the target of high-resolution spectroscopic observations of future missions, such as MUSE \citep{cheung2022probing, de2022probing} and SOLAR-C/EUVST \citep{shimizu2020solar}. Specific diagnostics will be the subject of a forthcoming investigation.

\begin{acknowledgements}
      \\
      GC, PP, and FR acknowledge support from ASI/INAF agreement n. 2022-29-HH.0 and from italian Ministero dell'Universit\'a e della Ricerca (MUR).
      \\
      JR, AWH and GC gratefully acknowledge the financial support of the Science and Technology Facilities Council (STFC) through Consolidated Grants ST/S000402/1 and ST/W001195/1 to the University of St Andrews.
      \\
      This work made use of the HPC system MEUSA, part of the Sistema Computazionale per l'Astrofisica Numerica (SCAN) of INAF-Osservatorio Astronomico di Palermo.
\end{acknowledgements}

%
%

\bibliography{bibliography}

\end{document}